\documentclass[11pt,a4paper]{article}

\usepackage[authoryear]{natbib}
\usepackage{graphicx}
\usepackage{amsmath}
\usepackage{amsfonts}
\usepackage{amssymb}
\usepackage[T1]{fontenc}
\usepackage{stix}
\usepackage{bm}
\usepackage{mathtools}
\usepackage{booktabs}
\usepackage{float}
\usepackage{placeins}
\usepackage{enumitem}
\usepackage{needspace}
\usepackage{indentfirst}
\usepackage{etoolbox}
\usepackage{microtype}
\usepackage[a4paper,margin=25mm]{geometry}
\usepackage[font=small,labelfont=bf,labelsep=colon]{caption}
\usepackage{hyperref}
\hypersetup{hypertexnames=false,hidelinks,
  pdftitle={A Higher-Order Stroh Framework for Weakly Nonlinear Post-Bifurcation Analysis},
  pdfauthor={Shiyuan Deng, Han Ye, Yafei Wang, Lizichen Chen, Alain Goriely, Yibin Fu, Weiqiu Chen, Chaofeng Lu, Michel Destrade, Kecheng Li, Yangkun Du}}

\graphicspath{{figures/}}
\ifdefined\pdfsuppresswarningpagegroup
\fi
\AtBeginDocument{}

\begin{document}

\begin{center}
{\LARGE\bfseries A Higher-Order Stroh Framework for Weakly Nonlinear Post-Bifurcation Analysis\par}
\vspace{1.2em}
{\large Shiyuan~Deng\textsuperscript{$a$}, Han~Ye\textsuperscript{$d$}, Yafei~Wang\textsuperscript{$c$}, Lizichen~Chen\textsuperscript{$a$}, Alain~Goriely\textsuperscript{$f$}, Yibin~Fu\textsuperscript{$g$}, Weiqiu~Chen\textsuperscript{$a,h$}, Chaofeng~L\"u\textsuperscript{$d,e,b,\ast$}, Michel~Destrade\textsuperscript{$h,\ast$}, Kecheng~Li\textsuperscript{$d,e,\ast$} and Yangkun~Du\textsuperscript{$a,\ast$}\par}
\vspace{1em}
{\small\itshape \textsuperscript{a}State Key Laboratory of Fluid Power and Mechatronic Systems, Department of Engineering Mechanics, Zhejiang University, Hangzhou 310027, China\\\textsuperscript{b}College of Civil Engineering and Architecture, Zhejiang University, Hangzhou 310058, China\\\textsuperscript{c}Center for Composite Materials and Structure, Harbin Institute of Technology, Harbin 150001, China\\\textsuperscript{d}Center for Mechanics Plus Under Extreme Environments, Ningbo University, Ningbo 315211, China\\\textsuperscript{e}Key Laboratory of Impact and Safety Engineering, Ministry of Education, Ningbo University, Ningbo 315211, China\\\textsuperscript{f}Mathematical Institute, University of Oxford, Oxford OX2 6GG, United Kingdom\\\textsuperscript{g}School of Computer Science and Mathematics, Keele University, Staffordshire ST5 5BG, United Kingdom\\\textsuperscript{h}School of Mathematical and Statistical Sciences, University of Galway, Galway, Ireland\par}
\end{center}

{\renewcommand{\thefootnote}{}\footnotetext{\raggedright$^\ast$Corresponding authors. E-mail addresses: \texttt{lucf@zju.edu.cn} (C. L\"u); \texttt{lucf@nbu.edu.cn} (C. L\"u); \texttt{michel.destrade@universityofgalway.ie} (M. Destrade); \texttt{likecheng@nbu.edu.cn} (K. Li); \texttt{yangkundu@zju.edu.cn} (Y. Du).}}

\begin{abstract}
Linear stability analysis predicts the onset of wrinkling, while weakly nonlinear analysis determines whether the bifurcation is supercritical or subcritical through the sign of the cubic amplitude coefficient. When this coefficient vanishes, higher-order terms are needed. In this paper, we develop a higher-order Stroh framework for weakly nonlinear post-bifurcation analysis of finitely deformed elastic solids. For each nonzero harmonic, the same base-state Stroh operator is retained at every order, with nonlinear source terms assembled from lower-order fields to account for geometry, constitutive laws and constraints. This recursive structure, together with Fredholm solvability conditions, enables us to derive cubic, quintic and higher-order amplitude equations. Here, we demonstrate the framework by analysing wrinkling in an incompressible neo-Hookean film bonded to an exponentially graded substrate. The results recover established linear instability benchmarks and the supercritical--subcritical transition in the homogeneous-substrate limit. For graded substrates, the cubic coefficient vanishes on a surface in the parameter space of modulus ratio and substrate gradation. On this surface, the quintic coefficient takes both signs, corresponding to supercritical and subcritical bifurcations. These regions are separated by a curve along which the quintic coefficient also vanishes. Determining the branch direction along this curve requires higher-order terms, which can be calculated within the same framework.
\end{abstract}

\noindent\textbf{Keywords:} higher-order Stroh formalism; weakly nonlinear post-bifurcation; surface wrinkling; graded substrate; cubic degeneracy; Fredholm solvability

\bigskip
\emergencystretch=2em

\setlength{\abovedisplayskip}{5pt plus 1pt minus 1pt}
\setlength{\belowdisplayskip}{5pt plus 1pt minus 1pt}
\setlength{\abovedisplayshortskip}{3pt plus 1pt minus 1pt}
\setlength{\belowdisplayshortskip}{3pt plus 1pt minus 1pt}
\setlength{\parindent}{1.5em}
\setlength{\parskip}{0pt}

\setcounter{section}{0}
\setcounter{figure}{0}
\setcounter{equation}{0}
\renewcommand{\theequation}{\arabic{equation}}
\renewcommand{\theHequation}{M\arabic{equation}}

\section{Introduction}\label{sec:introduction}

\begin{figure}[htbp]
\centering
\includegraphics[width=0.98\textwidth]{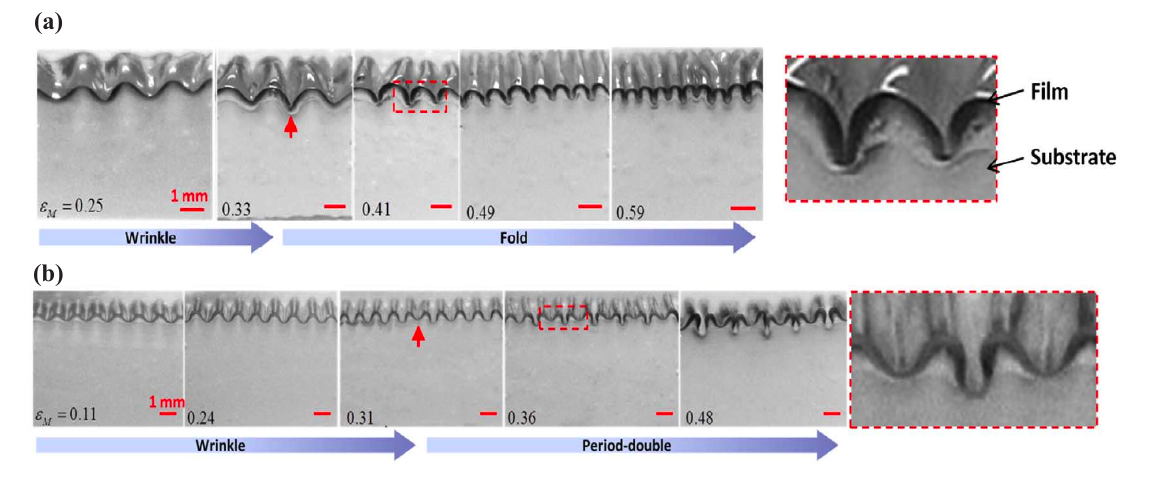}
\caption{Representative post-bifurcation morphologies in film--substrate systems. (a) Periodic wrinkles transition to localized folds as the mismatch strain \(\varepsilon_M\) increases, for \(\mu_f/\mu_s=3.64\). (b) A period-doubled pattern develops from wrinkles for \(\mu_f/\mu_s=67.24\) \citep{wangzhao2015}. Arrows indicate the emerging secondary morphology, and dashed boxes mark the enlarged regions.}
\label{fig:figure1}
\end{figure}

Surface instabilities are ubiquitous in soft matter and deformable surfaces, and are encountered in soft coatings, stretchable electronic devices, swelling systems, and growing biological tissues and soft materials \citep{bowden1998,stafford2004,rogers2010,li2012,tallinen2016,du2019,du2020,wang2024,su2024,qiu2025,liuliu2026}. When a system is subjected to compression, whether induced by mechanical loading, growth, prestretching, or differential swelling, surface wrinkling can occur once a critical condition is reached. A canonical example is a stiff thin film bonded to a softer substrate, where the stiffness contrast between the film and substrate determines the characteristic wavelength of the wrinkles \citep{cerda2003,huang2005}. When the interface is strongly bonded, increasing mismatch strain can drive the regular wrinkles into localized folds, period-doubling, and other post-buckling patterns (Fig.~\ref{fig:figure1}; \citealp{wangzhao2015}). These finite-amplitude morphologies are governed by higher-order nonlinearities and mode competition away from the critical point. The emerging branch is either supercritical or subcritical, and this determines the post-bifurcation behaviour. For a homogeneous half-space, Biot's linear instability theory provides the classical reference \citep{biot1963}, but linear bifurcation analysis determines only the critical load, wavelength, and mode, and cannot determine the direction of the bifurcating branch.

Near the critical point, weakly nonlinear analysis takes the linear critical bifurcation mode as the leading-order term and expands the incremental fields in powers of a small amplitude, with the fields at each order determined together with their corresponding boundary conditions \citep{green1952,budiansky1974}. At third order, the virtual-work formulation gives a solvability condition for the resonant fundamental harmonic and hence the cubic amplitude equation \citep{fuogden1999}. When the cubic coefficient is nonzero, its sign determines the local direction of the bifurcating branch. \citet{caifu1999} calculated this coefficient for a coated neo-Hookean half-space and found that it changes sign at a substrate-to-film shear modulus ratio of approximately 0.57, marking the transition between supercritical and subcritical bifurcation. Subsequent studies of film--substrate systems, using finite-deformation analyses, reduced plate--substrate models, and fully nonlinear bilayer or multilayer formulations, have further elucidated the roles of substrate nonlinear response, mode competition, and secondary pattern formation \citep{huang2005,audoly2008,brau2011,cao2012,hutchinson2013,alawiye2019,alawiye2020,zhou2022}. When the cubic coefficient vanishes, however, the cubic amplitude equation becomes degenerate and no longer determines the branch direction, requiring fifth-order or higher-order analysis. Although the third-order solvability condition is satisfied, the third-order resonant field is generally nonzero and reappears as a source term in the fifth-order problem. The mean field and higher harmonics are coupled progressively through nonlinear terms, so that the higher-order coefficients depend on a sequence of inhomogeneous field problems driven by the lower-order solutions.

The problem becomes more challenging for layered or continuously inhomogeneous substrates. The coefficients of the incremental equations vary with depth, and the depth-dependent modes generally have no closed-form expressions, so even the linear critical problem must be solved numerically. For two-dimensional elasticity, the Stroh formulation \citep{stroh1958,chadwick1997} introduces a state vector composed of displacement and traction variables and converts the second-order equilibrium equations in the depth direction into a first-order system of ordinary differential equations. For modes that vary periodically along the horizontal direction, the coefficients of this system depend only on depth. The decaying modes in the substrate can then be represented by a surface impedance matrix satisfying a Riccati equation \citep{fu2005,fu2007}, whose integration through the depth provides an efficient treatment of the variable coefficients. Such approaches have been used for several linear instability problems \citep{liu2024,wang2023}. Weakly nonlinear amplitude equations have also been derived for selected instability problems using virtual-work and semi-analytical formulations \citep{fuogden1999,caifu1999,jin2019}.

In this work, we develop a higher-order weakly nonlinear post-bifurcation framework for finitely deformed hyperelastic solids. For each nonzero harmonic, the incremental fields at all perturbation orders are governed by the same Stroh operator evaluated at the corresponding harmonic wavenumber of the critical state. The nonlinear source terms are assembled recursively from the lower-order fields, while the mean field is determined separately from the equilibrium and constraint equations. By imposing the Fredholm solvability condition at each resonant order, the framework yields the cubic, quintic, and, in principle, higher-order amplitude equations successively. We apply the framework to a compressed neo-Hookean film bonded to an exponentially graded neo-Hookean substrate. The calculated results recover existing linear instability benchmarks and, in the homogeneous-substrate limit, reproduce the transition between supercritical and subcritical bifurcation at a substrate-to-film shear modulus ratio of approximately 0.571. In the parameter space spanned by the gradation parameters and modulus ratio, the states at which the cubic coefficient vanishes form a cubic-degeneracy surface. On this surface, the quintic coefficient can be either positive or negative, corresponding to opposite branch directions. The two regions are separated by a curve on which the quintic coefficient also vanishes; the branch direction along this curve can only be determined by even higher-order analysis.

The rest of this paper is organized as follows. Sections 2 and 3 derive the nonlinear incremental equations at successive orders and formulate the recursive higher-order Stroh systems for incompressible and compressible materials. In Section 4, we derive the third- and fifth-order solvability conditions for the general film--substrate system. In Section 5, the film--substrate system is specialized to incompressible neo-Hookean materials with an exponentially graded substrate. Section 6 discusses the main results and the scope of the framework.

\section{Nonlinear incremental formulation}\label{nonlinear-incremental-formulation}

\begin{figure}[htbp]
\centering
\includegraphics[width=0.850\textwidth]{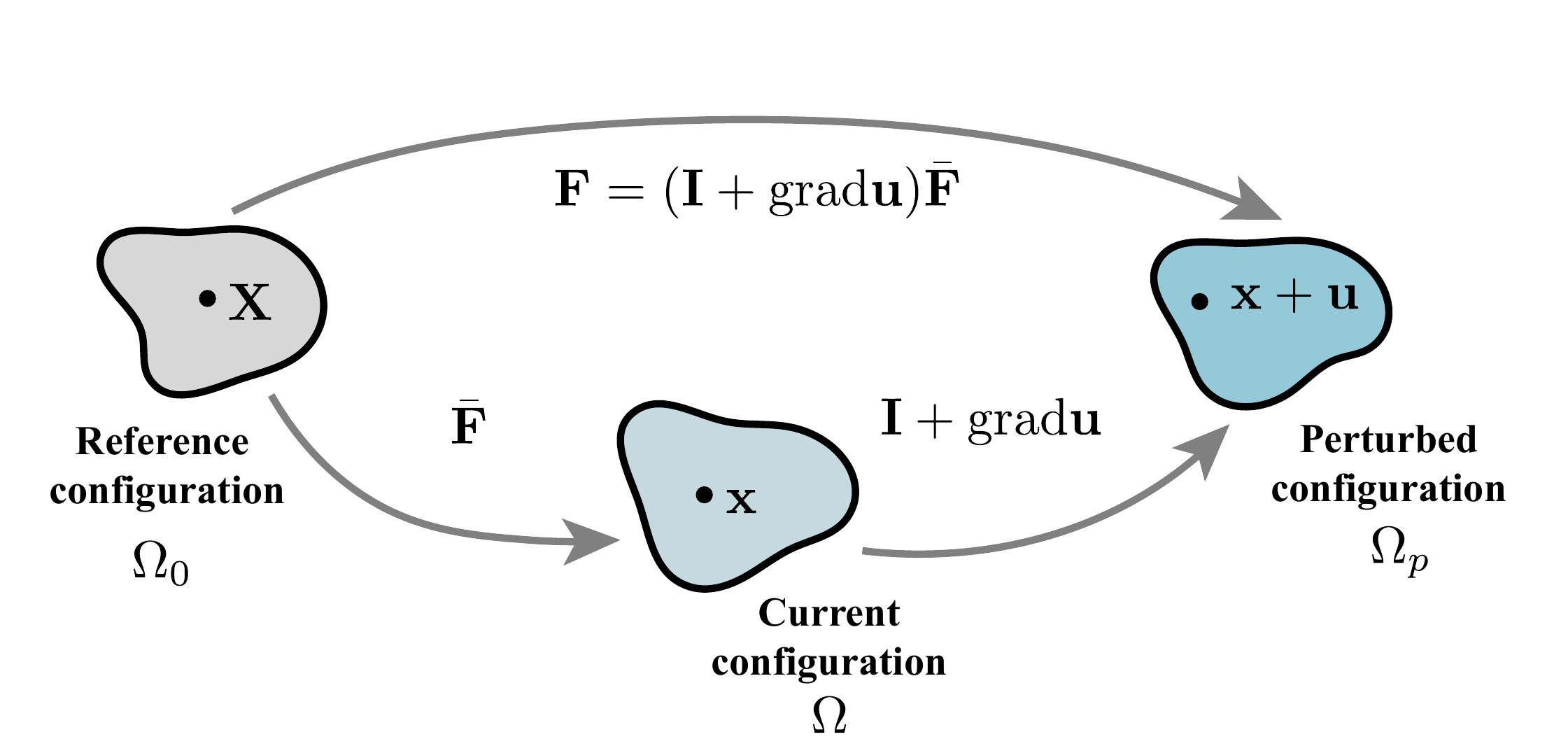}
\caption{Kinematic decomposition for the nonlinear incremental formulation. The finite deformation \(\bar{\mathbf F}\) maps the reference configuration \(\Omega_0\), with material point \(\mathbf X\), to the current base configuration \(\Omega\), with point \(\mathbf x\). Superposition of \(\mathbf u(\mathbf x,t)\) then gives the perturbed configuration \(\Omega_p\), with point \(\mathbf x+\mathbf u\). The composed mapping from the reference to the perturbed configuration is defined by \(\mathbf F=(\mathbf I+\operatorname{grad}\mathbf u)\bar{\mathbf F}\), with \(\operatorname{grad}\mathbf u\) taken in the current configuration.}
\label{fig:figure2}
\end{figure}

We consider a hyperelastic body undergoing a finite deformation from a reference configuration \(\Omega_0\), with material coordinates \(\mathbf X\), to a current configuration \(\Omega\), with spatial coordinates \(\mathbf x\). The base-state deformation gradient is denoted by \(\bar{\mathbf F}\), with \(\bar F_{iA}=\partial x_i/\partial X_A\), where capital and lower-case indices refer to the reference and current configurations, respectively. An incremental displacement \(\mathbf u=\mathbf u(\mathbf x,t)\) is superposed on the current configuration. The total deformation gradient is
\begin{equation}
\label{eq:main-1}
\mathbf F=(\mathbf I+\operatorname{grad}\mathbf u)\,\bar{\mathbf F},
\end{equation}
where \((\operatorname{grad}\mathbf u)_{ij}=u_{i,j}=\partial u_i/\partial x_j\). For compactness, we write \(\mathbf G:=\operatorname{grad}\mathbf u\) for the incremental displacement gradient. Repeated indices imply Einstein summation. Throughout this paper, we consider two-dimensional incremental motions in the \((x_1,x_2)\)-plane. The three configurations and their deformation mappings are illustrated in Fig.~\ref{fig:figure2}.

Let \(W(\mathbf F;\mathbf X)\) denote the strain-energy density per unit reference volume \citep{destradezurlo2025}. Its explicit dependence on \(\mathbf X\) allows for material inhomogeneity, and all constitutive derivatives with respect to \(\mathbf F\) are taken at fixed material coordinate \(\mathbf X\). We consider an incompressible material satisfying \(\det\mathbf F=1\). Introducing the Lagrange multiplier \(p\), the nominal stress is \(S_{iA}=\partial W/\partial F_{iA}-p(F^{-\mathrm{T}})_{iA}\). To express the incremental equations in the current configuration, we define the push-forward incremental nominal stress by
\begin{equation}
\label{eq:main-2}
\chi_{ij} = \bar J^{-1} \bigl(S_{iA}-\bar S_{iA}\bigr)\bar F_{jA}, \qquad \bar J=\det\bar{\mathbf F},
\end{equation}
where \(\bar{\mathbf S}\) is the nominal stress in the base state. For the incompressible base state, \(\bar J=1\). Expanding Eq. \eqref{eq:main-2} through cubic order in the incremental fields \(\mathbf G\) and \(\tilde p\) gives
\begin{equation}
\label{eq:main-3}
\begin{aligned}
\chi_{ij}
=
&\,
\mathcal A^1_{jilk}u_{k,l}
+\bar p\,u_{j,i}
-\tilde p\,\delta_{ij}
+
\frac12\mathcal A^2_{jilk\,nm}u_{k,l}u_{m,n}
+
\frac16\mathcal A^3_{jilk\,nm\,qp}
u_{k,l}u_{m,n}u_{p,q}
\\
&+
\bar p
\left(
-u_{j,k}u_{k,i}
+
u_{j,k}u_{k,l}u_{l,i}
\right)
+
\tilde p
\left(
u_{j,i}
-
u_{j,k}u_{k,i}
\right)
+
\text{higher-order terms}.
\end{aligned}
\end{equation}
Here \(p=\bar p+\tilde p\), where \(\bar p\) is the base-state pressure and \(\tilde p\) is its increment. The instantaneous material tensors are defined by
\begin{equation}
\label{eq:main-4}
\begin{aligned}
&\mathcal A^1_{jilk}
=
\bar J^{-1}\bar F_{jA}\bar F_{lB}
\frac{\partial^2W}
{\partial F_{iA}\partial F_{kB}}
\bigg|_{\bar{\mathbf F}},\quad
\mathcal A^2_{jilk\,nm}
=
\bar J^{-1}\bar F_{jA}\bar F_{lB}\bar F_{nC}
\frac{\partial^3W}
{\partial F_{iA}\partial F_{kB}\partial F_{mC}}
\bigg|_{\bar{\mathbf F}},
\\
&\mathcal A^3_{jilk\,nm\,qp}
=
\bar J^{-1}\bar F_{jA}\bar F_{lB}\bar F_{nC}\bar F_{qD}
\frac{\partial^4W}
{\partial F_{iA}\partial F_{kB}\partial F_{mC}\partial F_{pD}}
\bigg|_{\bar{\mathbf F}} .
\end{aligned}
\end{equation}

The tensor \(\boldsymbol{\mathcal A}^1\) gives the strain-energy contribution to the linear incremental response, while \(\boldsymbol{\mathcal A}^2\) and \(\boldsymbol{\mathcal A}^3\) enter the quadratic and cubic constitutive contributions, respectively. The term \(\bar p\,u_{j,i}\) in Eq. \eqref{eq:main-3} is also linear in the incremental displacement gradient and is therefore incorporated into the effective first-order tensor \(\bar{\mathcal A}^{1}_{jilk}=\mathcal A^1_{jilk}+\bar p\,\delta_{jk}\delta_{il}\). The incremental stress can then be written as
\begin{equation}
\label{eq:main-5}
\chi_{ij} = \bar{\mathcal A}^{1}_{jilk}u_{k,l} - \tilde p\,\delta_{ij} + \mathcal S_{ij},
\end{equation}
where the nonlinear contributions are collected in the stress-source tensor \(\boldsymbol{\mathcal S}\),
\begin{equation}
\label{eq:main-6}
\begin{split}\mathcal S_{ij} ={}& \frac12\mathcal A^2_{jilk\,nm}u_{k,l}u_{m,n} + \frac16\mathcal A^3_{jilk\,nm\,qp} u_{k,l}u_{m,n}u_{p,q}\\ &+ \bar p \left( -u_{j,k}u_{k,i} + u_{j,k}u_{k,l}u_{l,i} \right) + \tilde p \left( u_{j,i} - u_{j,k}u_{k,i} \right) + \text{higher-order terms}.\end{split}
\end{equation}

In the absence of body forces, the incremental equations of motion are
\begin{equation}
\label{eq:main-7}
\chi_{ij,j} = \rho\frac{\partial^2u_i}{\partial t^2}, \qquad i=1,2,
\end{equation}
where \(\rho\) is the mass density per unit current volume.

The incompressibility constraint can be separated in the same way. Since \(\det\bar{\mathbf F}=1\), incompressibility of the total deformation requires \(\det(\mathbf I+\mathbf G)=1\). For two-dimensional incremental motions, this condition can be written exactly as
\begin{equation}
\label{eq:main-8}
u_{i,i} = \mathcal D, \qquad \mathcal D = -\frac12 \left( u_{i,i}u_{j,j} - u_{i,j}u_{j,i} \right).
\end{equation}
Here \(u_{i,i}\) is the linear part of the incompressibility condition, while \(\mathcal D\) collects its nonlinear contribution. In two dimensions, Eq. \eqref{eq:main-8} is exact because \(\det(\mathbf I+\mathbf G)\) is exactly quadratic in the entries of the \(2\times2\) matrix \(\mathbf G\). Although \(\mathcal D\) is only quadratic in the total gradient \(\mathbf G\), substituting a perturbation expansion for \(\mathbf G\) generates contributions at every order \(n\ge2\) through products of lower-order fields.

\section{Higher-order Stroh framework}\label{higher-order-stroh-framework}

The classical Stroh formalism recasts the linear incremental equations of a finitely deformed elastic solid as a first-order system in one spatial coordinate. We extend this structure to the weakly nonlinear problem by organizing the perturbation equations recursively: at every order, the same linear Stroh operator associated with the base state acts on the unknown field, while nonlinear interactions among previously determined fields enter as known forcing. This order-independent operator structure is the central feature of the higher-order formulation.

\subsection{Recursive perturbation hierarchy}\label{recursive-perturbation-hierarchy}

We introduce a small amplitude parameter \(\varepsilon\) and expand the incremental displacement, pressure, stress, and displacement gradient as
\begin{equation}
\label{eq:main-9}
\mathbf u = \sum_{n\geq1}\varepsilon^n\mathbf u^{(n)}, \tilde p = \sum_{n\geq1}\varepsilon^n p^{(n)}, \qquad \boldsymbol\chi = \sum_{n\geq1}\varepsilon^n\boldsymbol\chi^{(n)}, \mathbf G = \sum_{n\geq1}\varepsilon^n\mathbf G^{(n)}, \qquad \mathbf G^{(n)}=\operatorname{grad}\mathbf u^{(n)} .
\end{equation}
Substituting these expansions into the governing equations of Section 2 and collecting terms of order \(\varepsilon^n\) gives
\begin{equation}
\label{eq:main-10}
\chi^{(n)}_{ij} = \bar{\mathcal A}^{1}_{jilk}u^{(n)}_{k,l} - p^{(n)}\delta_{ij} + \mathcal S^{(n)}_{ij}, \qquad \chi^{(n)}_{ij,j} = \rho\frac{\partial^2u^{(n)}_i}{\partial t^2}, \qquad u^{(n)}_{i,i} = \mathcal D^{(n)}.
\end{equation}
Here \(\mathcal S^{(n)}_{ij}\) and \(\mathcal D^{(n)}\) are the order-\(n\) nonlinear stress and incompressibility sources. After the linear terms involving \(\mathbf u^{(n)}\) and \(p^{(n)}\) have been separated as in Eq. \eqref{eq:main-10}, both sources depend only on fields of order lower than \(n\). Thus, once the lower-order fields are known, the order-\(n\) problem is linear in the new unknowns.

For the two-dimensional incompressibility condition, the source is generated exactly by
\begin{equation}
\label{eq:main-11}
\mathcal D^{(n)} = -\frac12 \sum_{a=1}^{n-1} \left[ G^{(a)}_{ii}G^{(n-a)}_{jj} - G^{(a)}_{ij}G^{(n-a)}_{ji} \right], \qquad n\geq2.
\end{equation}
Equation \eqref{eq:main-11} generates the higher-order constraint sources through products of lower-order gradients. At first order, \(\mathcal S^{(1)}=\mathcal D^{(1)}=0\). For \(n\ge2\), the stress source \(\mathcal S^{(n)}\) is obtained by the same order-collection procedure from the nonlinear constitutive expansion. Its arbitrary-order construction and the explicit fourth- and fifth-order expressions are given in Appendix~\ref{higher-order-nonlinear-sources}.

\subsection{Forced Stroh system for incompressible materials}\label{forced-stroh-system-for-incompressible-materials}

We recast the order-\(n\) problem in Stroh form. From the effective first-order instantaneous tensor \(\bar{\boldsymbol{\mathcal A}}^1\), define
\begin{equation}
\label{eq:main-12}
\mathbf Q=[\bar{\mathcal A}^{1}_{1i1k}], \qquad \mathbf R=[\bar{\mathcal A}^{1}_{1i2k}], \qquad \mathbf T=[\bar{\mathcal A}^{1}_{2i2k}].
\end{equation}
We group the stress and nonlinear source components as \(\boldsymbol\chi^{(n)}_1=(\chi^{(n)}_{11},\chi^{(n)}_{21})^{\mathrm{T}}\), \(\boldsymbol\chi^{(n)}_2=(\chi^{(n)}_{12},\chi^{(n)}_{22})^{\mathrm{T}}\), \(\mathbf s^{(n)}_1=(\mathcal S^{(n)}_{11},\mathcal S^{(n)}_{21})^{\mathrm{T}}\), and \(\mathbf s^{(n)}_2=(\mathcal S^{(n)}_{12},\mathcal S^{(n)}_{22})^{\mathrm{T}}\). Equation \eqref{eq:main-10} then gives
\begin{equation}
\label{eq:main-13}
\boldsymbol\chi^{(n)}_1 = \mathbf Q\,\mathbf u^{(n)}_{,1} + \mathbf R\,\mathbf u^{(n)}_{,2} - p^{(n)}\mathbf e_1 + \mathbf s^{(n)}_1, \qquad \boldsymbol\chi^{(n)}_2 = \mathbf R^{\mathrm{T}}\mathbf u^{(n)}_{,1} + \mathbf T\,\mathbf u^{(n)}_{,2} - p^{(n)}\mathbf e_2 + \mathbf s^{(n)}_2,
\end{equation}
where \(\mathbf e_1=(1,0)^{\mathrm{T}}\) and \(\mathbf e_2=(0,1)^{\mathrm{T}}\). For fields with non-zero variation in \(x_1\), introduce a stress-potential vector \(\boldsymbol\phi^{(n)}\) and an \(x_1\)-antiderivative \(\mathbf w^{(n)}\) of the displacement,
\begin{equation}
\label{eq:main-14}
\boldsymbol\chi^{(n)}_2 = \boldsymbol\phi^{(n)}_{,1}, \qquad \mathbf w^{(n)}_{,1} = \mathbf u^{(n)}.
\end{equation}
The equilibrium equations then imply
\begin{equation}
\label{eq:main-15}
\boldsymbol\chi^{(n)}_1 = -\boldsymbol\phi^{(n)}_{,2} + \rho\frac{\partial^2\mathbf w^{(n)}}{\partial t^2}.
\end{equation}

The representation in Eq. \eqref{eq:main-14} requires non-zero variation along \(x_1\). Accordingly, any zero-wavenumber mean field generated by nonlinear interactions is solved directly from the original equilibrium equations, whereas the Stroh reduction below applies to the non-zero harmonics. For these harmonics, we collect the displacement and stress potential into the Stroh state
\begin{equation}
\label{eq:main-16}
\boldsymbol\eta^{(n)} = \begin{pmatrix} \mathbf u^{(n)}, \boldsymbol\phi^{(n)} \end{pmatrix}^{\mathrm{T}}.
\end{equation}

For incompressible materials, \(p^{(n)}\) is an additional unknown and must be eliminated together with the incompressibility constraint. We write the latter in residual form as
\begin{equation}
\label{eq:main-17}
\mathcal C^{(n)} = \mathbf e_1^{\mathrm{T}}\mathbf u^{(n)}_{,1} + \mathbf e_2^{\mathrm{T}}\mathbf u^{(n)}_{,2} - \mathcal D^{(n)} = 0.
\end{equation}
Adding multiples of the vanishing residual \(\mathcal C^{(n)}\), introduce \(\kappa>0\) and define
\begin{equation}
\label{eq:main-18}
\mathbf Q_\kappa = \mathbf Q+\kappa\,\mathbf e_1\otimes\mathbf e_1, \qquad \mathbf R_\kappa = \mathbf R+\kappa\,\mathbf e_1\otimes\mathbf e_2, \qquad \mathbf T_\kappa = \mathbf T+\kappa\,\mathbf e_2\otimes\mathbf e_2,
\end{equation}
with \(\kappa\) chosen so that \(\mathbf T_\kappa\) is invertible. For compactness, introduce the auxiliary quantities
\begin{equation}
\label{eq:stroh-auxiliary-quantities}
\zeta_\kappa = \mathbf e_2^{\mathrm{T}} \mathbf T_\kappa^{-1} \mathbf e_2, \qquad \mathbf b_\kappa^{\mathrm{T}} = \mathbf e_2^{\mathrm{T}} \mathbf T_\kappa^{-1} \mathbf R_\kappa^{\mathrm{T}} - \mathbf e_1^{\mathrm{T}}.
\end{equation}

Eliminating \(p^{(n)}\) using Eq. \eqref{eq:main-17} then yields the forced Stroh system
\begin{equation}
\label{eq:main-19}
\frac{\partial\boldsymbol\eta^{(n)}}{\partial x_2} = \mathbf N \frac{\partial\boldsymbol\eta^{(n)}}{\partial x_1} + \mathbf H^{(n)} + \rho\frac{\partial^2}{\partial t^2} \begin{pmatrix} \mathbf 0 \\ \mathbf w^{(n)} \end{pmatrix},
\end{equation}
where the Stroh matrix and its blocks are given together by
\begin{equation}
\label{eq:main-20}
\begin{aligned}
&\mathbf N
=
\begin{pmatrix}
\mathbf N_1 & \mathbf N_2 \\
\mathbf N_3 & \mathbf N_1^{\mathrm{T}}
\end{pmatrix},
\\[2pt]
&\mathbf N_1
=
-\mathbf T_\kappa^{-1}\mathbf R_\kappa^{\mathrm{T}}
+
\zeta_\kappa^{-1}
\left(
\mathbf T_\kappa^{-1}\mathbf e_2
\right)
\mathbf b_\kappa^{\mathrm{T}},
\\
&\mathbf N_2
=
\mathbf T_\kappa^{-1}
-
\zeta_\kappa^{-1}
\left(
\mathbf T_\kappa^{-1}\mathbf e_2
\right)
\left(
\mathbf e_2^{\mathrm{T}}\mathbf T_\kappa^{-1}
\right),
\\
&\mathbf N_3
=
\mathbf R_\kappa
\mathbf T_\kappa^{-1}
\mathbf R_\kappa^{\mathrm{T}}
-
\mathbf Q_\kappa
+
\zeta_\kappa^{-1}
\left(
\mathbf e_1
-
\mathbf R_\kappa
\mathbf T_\kappa^{-1}\mathbf e_2
\right)
\mathbf b_\kappa^{\mathrm{T}},
\end{aligned}
\end{equation}
and the nonlinear source vector \(\mathbf H^{(n)}=(\mathbf H_u^{(n)},\mathbf H_\phi^{(n)})^{\mathrm{T}}\) is
\begin{equation}
\label{eq:main-23}
\mathbf H_u^{(n)} = -\mathbf N_2\mathbf s_2^{(n)} + \zeta_\kappa^{-1} \mathbf T_\kappa^{-1}\mathbf e_2\, \mathcal D^{(n)}, \qquad \mathbf H_\phi^{(n)} = -\mathbf s_1^{(n)} - \mathbf N_1^{\mathrm{T}}\mathbf s_2^{(n)} - \zeta_\kappa^{-1} \mathbf b_\kappa\, \mathcal D^{(n)}.
\end{equation}
The assembled \(\mathbf N\) and \(\mathbf H^{(n)}\) are independent of \(\kappa\); the elimination is given in Appendix~\ref{constrained-stroh-reduction-and-fredholm-projection}.

Defining the principal spatial Stroh operator as \(\mathcal L_\eta:=\partial/\partial x_2-\mathbf N\,\partial/\partial x_1\), Eq. \eqref{eq:main-19} becomes
\begin{equation}
\label{eq:main-24}
 \mathcal L_\eta\boldsymbol\eta^{(n)} = \mathbf H^{(n)} + \rho\frac{\partial^2}{\partial t^2} \begin{pmatrix} \mathbf 0 \\ \mathbf w^{(n)} \end{pmatrix}.
\end{equation}
Equation \eqref{eq:main-24} is the higher-order Stroh hierarchy: the base-state operator is common to all orders, while \(\mathbf H^{(n)}\) supplies the order-dependent forcing.
At first order, \(\mathcal S^{(1)}=\mathcal D^{(1)}=0\), so \(\mathbf H^{(1)}=\mathbf0\). For a time-independent bifurcation mode, Eq. \eqref{eq:main-24} therefore reduces to
\begin{equation}
\label{eq:main-25}
 \mathcal L_\eta\boldsymbol\eta^{(1)} =\mathbf 0,
 \frac{\partial\boldsymbol\eta^{(1)}}{\partial x_2} = \mathbf N \frac{\partial\boldsymbol\eta^{(1)}}{\partial x_1},
\end{equation}
which recovers the classical linear incremental Stroh system.

\subsection{Compressible counterpart}\label{compressible-counterpart}

The same recursive structure is not restricted to incompressibility. For a compressible material, the pressure field and incompressibility constraint are absent, and the order-\(n\) constitutive relation is
\begin{equation}
\label{eq:main-26}
\chi^{(n)}_{ij} = \mathcal A^{1}_{jilk}u^{(n)}_{k,l} + \mathcal S^{(n)}_{ij}.
\end{equation}
The matrices \(\mathbf Q\), \(\mathbf R\), and \(\mathbf T\) are then constructed directly from \(\boldsymbol{\mathcal A}^{1}\). Under strong ellipticity, \(\mathbf T\) is positive definite and may be inverted directly, giving
\begin{equation}
\label{eq:main-27}
\mathbf N_1 = -\mathbf T^{-1}\mathbf R^{\mathrm{T}}, \mathbf N_2 = \mathbf T^{-1}, \mathbf N_3 = \mathbf R\mathbf T^{-1}\mathbf R^{\mathrm{T}} - \mathbf Q, \qquad \mathbf H_u^{(n)} = -\mathbf T^{-1}\mathbf s_2^{(n)}, \mathbf H_\phi^{(n)} = \mathbf R\mathbf T^{-1}\mathbf s_2^{(n)} - \mathbf s_1^{(n)}.
\end{equation}
The compressible problem therefore retains the same order-independent operator/recursive-forcing structure.

\section{Weakly nonlinear analysis of a film--substrate system}\label{weakly-nonlinear-analysis-of-a-filmsubstrate-system}

We apply the higher-order Stroh framework to the weakly nonlinear wrinkling of a compressed film bonded to an elastic substrate whose properties may vary with depth. We first reduce the recursive hierarchy into spatial harmonics and determine the linear critical state through the surface-impedance formulation. The resulting critical mode is then used to compute the forced second-order fields and the third-order Fredholm condition, which gives the cubic amplitude equation. When the cubic coefficient vanishes, the hierarchy is continued to fifth order to obtain the next non-zero nonlinear contribution.

\subsection{Film--substrate system and linear critical state}\label{filmsubstrate-system-and-linear-critical-state}

\begin{figure}[htbp]
\centering
\includegraphics[width=0.86\textwidth]{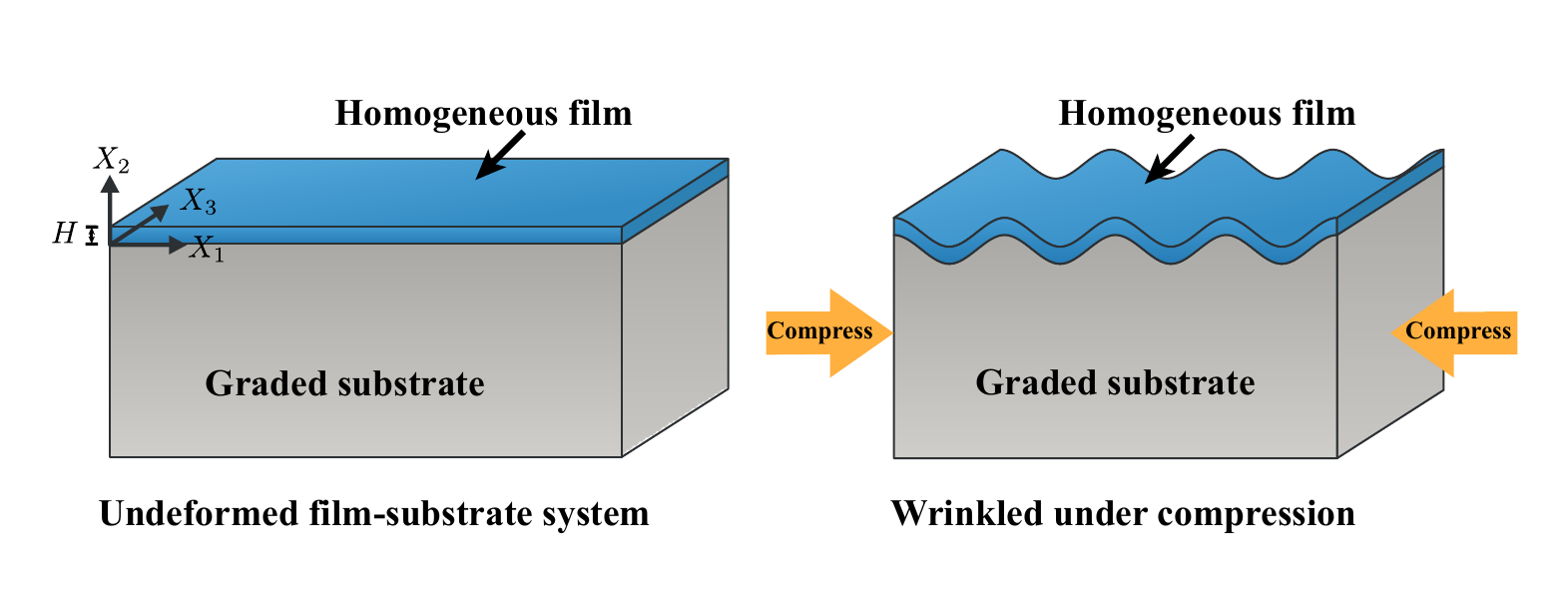}
\caption{Film--substrate configuration and compression-induced surface wrinkling. A homogeneous film of reference thickness \(H\) is perfectly bonded to a graded half-space with properties varying along the reference depth \(X_2\). Under uniform compression with stretch \(\lambda\) along \(X_1\), the system becomes unstable and develops periodic surface wrinkles. The left and right views show the reference and wrinkled configurations, respectively, with current film thickness \(h=H/\lambda\) and a traction-free upper surface.}
\label{fig:figure3}
\end{figure}

We consider a homogeneous film of reference thickness \(H\), perfectly bonded to a graded substrate occupying a half-space, as illustrated in Fig.~\ref{fig:figure3}. The material properties of the substrate vary continuously with the reference depth \(X_2\). The system is subjected to uniform compression with stretch \(\lambda\) along \(X_1\), leading to surface wrinkling beyond the critical state. The homogeneous base deformation is \(\bar{\mathbf F}(\lambda)=\operatorname{diag}(\lambda,\lambda^{-1},1)\), and, for an incompressible film, the current thickness is \(h=H/\lambda\). In the deformed configuration, the film occupies \(0\le x_2\le h\) and the substrate \(x_2<0\). The upper surface is traction free, displacement and traction are continuous across the interface \(x_2=0\), and the incremental fields decay as \(x_2\to-\infty\). Since the substrate grading is prescribed in the reference configuration, the homogeneous base deformation gives \(X_2=\lambda x_2\), so the instantaneous material tensors are uniform in the film but vary with depth in the substrate.

For fields periodic along \(x_1\), let \(m>0\) denote the fundamental wavenumber. Nonlinear interactions generate harmonics \(q=\ell m\), with \(\ell\in\mathbb Z\). At perturbation order \(n\), the Stroh state is written as
\begin{equation}
\label{eq:main-28}
\boldsymbol\eta^{(n)}(x_1,x_2,t) = \sum_{q\in\mathcal Q_n} \hat{\boldsymbol\eta}^{(n,q)}(x_2,t)e^{\mathrm iqx_1}, \qquad \hat{\boldsymbol\eta}^{(n,-q)} = \overline{\hat{\boldsymbol\eta}^{(n,q)}}.
\end{equation}
For a non-zero harmonic \(q\), the relation \(\mathbf w^{(n)}_{,1}=\mathbf u^{(n)}\) gives \(\hat{\mathbf w}^{(n,q)}=\hat{\mathbf u}^{(n,q)}/(\mathrm iq)\). Equation \eqref{eq:main-24} of Section 3 therefore reduces to
\begin{equation}
\label{eq:main-29}
\mathcal L_q(\lambda)\hat{\boldsymbol\eta}^{(n,q)} = \hat{\mathbf H}^{(n,q)} + \frac{\rho}{\mathrm iq} \frac{\partial^2}{\partial t^2} \begin{pmatrix} \mathbf0\\ \hat{\mathbf u}^{(n,q)} \end{pmatrix}, \qquad \mathcal L_q(\lambda) := \frac{d}{dx_2} - \mathrm iq\,\mathbf N(x_2;\lambda).
\end{equation}
Thus every non-zero harmonic is governed by the same Stroh matrix \(\mathbf N\), with the harmonic wavenumber entering only through \(q\). The zero-wavenumber mean field cannot be treated by this harmonic reduction and is solved directly from the original equilibrium and incompressibility equations.

Let \(\boldsymbol\xi\) denote the material and structural parameters that specify a particular film--substrate system. The loading stretch \(\lambda\), fundamental wavenumber \(m\), and response amplitude are not included in \(\boldsymbol\xi\). For each prescribed \(\boldsymbol\xi\), the linear problem determines a critical loading state and mode. Once a system is fixed, we write \(\lambda_c\), \(m_c\), and \(\hat{\boldsymbol\eta}_c\) instead of repeatedly writing \(\lambda_c(\boldsymbol\xi)\), \(m_c(\boldsymbol\xi)\), and \(\hat{\boldsymbol\eta}_c(\boldsymbol\xi)\). At first order, \(\mathbf H^{(1)}=\mathbf0\), and the static fundamental harmonic satisfies the homogeneous form of Eq. \eqref{eq:main-29}. For prescribed \(m\) and \(\lambda\), the film and substrate responses are obtained from the surface impedance matrix \(\mathbf Z\), which is constructed from the corresponding Stroh matrix \(\mathbf N\).

In the stress-potential convention adopted here, the impedance is defined by \(\hat{\boldsymbol\phi}^{(1,m)}=\mathbf Z\,\hat{\mathbf u}^{(1,m)}\). Since \(\hat{\boldsymbol\chi}^{(1,m)}_2=\mathrm im\,\hat{\boldsymbol\phi}^{(1,m)}\), this differs from the conventional traction-to-displacement impedance only by the common Fourier factor \(\mathrm im\). Differentiating the impedance relation and using the Stroh equations gives the Riccati equation
\begin{equation}
\label{eq:main-30}
\frac{d\mathbf Z}{dx_2} = \mathrm im \left( \mathbf N_3 + \mathbf N_1^{\mathrm{T}}\mathbf Z - \mathbf Z\mathbf N_1 - \mathbf Z\mathbf N_2\mathbf Z \right).
\end{equation}

In the semi-infinite substrate, the two independent depth modes that decay as \(x_2\rightarrow-\infty\) determine the substrate impedance, which is propagated to the interface to obtain \(\mathbf Z_{\mathrm{sub}}(0)\). In the homogeneous film, the traction-free condition at \(x_2=h\) gives \(\mathbf Z_{\mathrm{film}}(h)=\mathbf0\), and propagation through the film gives \(\mathbf Z_{\mathrm{film}}(0)\).
Perfect bonding requires continuity of displacement and traction across the interface. If \(\hat{\mathbf u}_0\) denotes the common non-zero interfacial displacement, then \(\mathbf Z_{\mathrm{film}}(0)\hat{\mathbf u}_0=\mathbf Z_{\mathrm{sub}}(0)\hat{\mathbf u}_0\). A non-trivial interfacial mode therefore exists when
\begin{equation}
\label{eq:main-31}
\det\!\left[ \mathbf Z_{\mathrm{film}}(0) - \mathbf Z_{\mathrm{sub}}(0) \right] =0.
\end{equation}

For each prescribed \(\boldsymbol\xi\) and \(m\), Eq. \eqref{eq:main-31} determines a bifurcation stretch \(\lambda_{\mathrm{bif}}(m;\boldsymbol\xi)\). Since increasing compression corresponds to decreasing \(\lambda\) from unity, the first instability encountered is the one with the largest bifurcation stretch,
\begin{equation}
\label{eq:main-32}
\lambda_c(\boldsymbol\xi) = \max_{m>0}\lambda_{\mathrm{bif}}(m;\boldsymbol\xi), \qquad m_c(\boldsymbol\xi) = \operatorname*{arg\,max}_{m>0}\lambda_{\mathrm{bif}}(m;\boldsymbol\xi).
\end{equation}
At \((\lambda_c,m_c)\), the interface matrix in Eq. \eqref{eq:main-31} is singular. Its null space determines the interfacial displacement, from which the complete depth-dependent critical state \(\hat{\boldsymbol\eta}_c(x_2)\) is reconstructed subject to the free-surface condition, interfacial continuity, and decay into the substrate.
Defining \(\mathbf N_c(x_2)=\mathbf N(x_2;\lambda_c)\), the normalized critical mode satisfies \(d\hat{\boldsymbol\eta}_c/dx_2=\mathrm i m_c\mathbf N_c\hat{\boldsymbol\eta}_c\). Its normalization is fixed and used throughout the higher-order calculation. We assume that the selected critical state is simple and isolated, with a one-dimensional critical eigenspace and no simultaneous degeneracy with another wavenumber or mode. Under this assumption, the critical branch and its modal fields vary smoothly with the system parameters in a sufficiently small neighbourhood.

For a fixed system, we define the compression detuning by
\begin{equation}
\label{eq:main-33}
\Delta:=\lambda_c(\boldsymbol\xi)-\lambda.
\end{equation}
Thus \(\boldsymbol\xi\) identifies the material/structural system, whereas \(\Delta\) measures loading relative to that system's own critical stretch. Positive \(\Delta\) corresponds to the post-critical compression side and negative \(\Delta\) to the pre-critical side.

\subsection{Cubic solvability and local branch direction}\label{cubic-solvability-and-local-branch-direction}

We expand about the critical state. The leading-order field contains the harmonics \(\pm m_c\). Their quadratic interaction generates only the mean and second harmonics, \(0\) and \(\pm2m_c\). The critical harmonic therefore first reappears in the nonlinear forcing at \(O(\varepsilon^3)\). With \(\boldsymbol\xi\) fixed, the detuning defined in Section 4.1 reduces to \(\Delta=\lambda_c-\lambda\). We call a branch \emph{supercritical} if it initially leaves the critical point toward \(\Delta>0\), and \emph{subcritical} if it initially leaves toward \(\Delta<0\).

Near onset, let \(\tau=\varepsilon^\sigma t\) and \(\Delta=O(\varepsilon^b)\). Because the leading perturbation is \(O(\varepsilon)\), slow-time inertia is \(O(\varepsilon^{1+2\sigma})\), while a loading departure of \(O(\varepsilon^b)\) produces an \(O(\varepsilon^{1+b})\) correction when the linear operator is expanded about \(\lambda_c\). Balancing both contributions with the first resonant nonlinear forcing at \(O(\varepsilon^3)\) gives \(\sigma=1\) and \(b=2\), hence
\begin{equation}
\label{eq:main-34}
\tau=\varepsilon t, \qquad \Delta=\varepsilon^2\Delta_2+O(\varepsilon^4).
\end{equation}
or equivalently \(\lambda=\lambda_c-\varepsilon^2\Delta_2+O(\varepsilon^4)\). The leading-order field is therefore written as
\begin{equation}
\label{eq:main-35}
\boldsymbol\eta^{(1)} = A(\tau)\hat{\boldsymbol\eta}_c(x_2)e^{\mathrm i m_cx_1} + \overline{A(\tau)}\,\overline{\hat{\boldsymbol\eta}_c(x_2)}e^{-\mathrm i m_cx_1}.
\end{equation}

The harmonic structure through third order is \(\mathcal Q_1=\{\pm m_c\}\), \(\mathcal Q_2=\{0,\pm2m_c\}\), and \(\mathcal Q_3=\{\pm m_c,\pm3m_c\}\). The fundamental harmonic reappears at third order through interactions such as \(0+m_c=m_c\) and \(2m_c-m_c=m_c\). This is the first order beyond the linear problem at which the singular critical operator is forced resonantly, so a nonlinear solvability condition is required.

For later use, recall that \(G_{ij}^{(n)}=u_{i,j}^{(n)}\). After Fourier expansion in \(x_1\), \(\hat G_{i1}^{(n,q)}=\mathrm iq\,\hat u_i^{(n,q)}\) and \(\hat G_{i2}^{(n,q)}=(\hat u_i^{(n,q)})'\). Thus the \(x_1\)-derivative becomes multiplication by \(\mathrm iq\), while the \(x_2\)-derivative remains a depth derivative. The symbols \(\hat{\boldsymbol\eta}_c\), \(\hat{\mathbf u}_c\), \(\hat G_{c,ij}\), and \(\hat p_c\) denote the normalized unit-amplitude critical eigenfields, so \(\hat{\boldsymbol\eta}^{(1,m_c)}=A\hat{\boldsymbol\eta}_c\) and \(\hat p^{(1,m_c)}=A\hat p_c\). At second order, the state has the form
\begin{equation}
\label{eq:main-second-order-state}
\boldsymbol\eta^{(2)} = |A|^2\hat{\boldsymbol\eta}^{(2,0)} + A^2\hat{\boldsymbol\eta}^{(2,2m_c)}e^{2\mathrm i m_cx_1} + \overline A^{\,2}\overline{\hat{\boldsymbol\eta}^{(2,2m_c)}}e^{-2\mathrm i m_cx_1}.
\end{equation}
The second harmonic is obtained directly from the forced Stroh system, \(\mathcal L_{2m_c,c}\hat{\boldsymbol\eta}^{(2,2m_c)}=\hat{\mathbf H}^{(2,2m_c)}\), where \(\mathcal L_{2m_c,c}=d/dx_2-2\mathrm i m_c\mathbf N_c\).

The mean field requires separate treatment. Since the stress-potential reduction used in Eq. \eqref{eq:main-29} relies on \(q\neq0\), the \(q=0\) component must instead be obtained from the unintegrated second-order equations,
\begin{equation}
\label{eq:main-36}
\chi_{ij}^{(2,0)} = \bar{\mathcal A}^{1}_{jilk}u_{k,l}^{(2,0)} - p^{(2,0)}\delta_{ij} + \mathcal S_{ij}^{(2,0)}, \qquad \chi_{ij,j}^{(2,0)}=0, \qquad u_{i,i}^{(2,0)}=\mathcal D^{(2,0)}.
\end{equation}
Both second-order fields satisfy the traction-free surface condition, continuity of displacement and traction at the interface, and decay conditions in the substrate. For the mean field, the far-field mean displacement is set to zero to fix the rigid-translation freedom.

At third order, the mean and second-harmonic corrections feed back into the fundamental. The resulting \(m_c\)-harmonic is governed by the singular operator \(\mathcal L_{m_c,c}\), so its forcing must satisfy a Fredholm condition. The nonlinear contribution to the Fredholm projection is
\begin{equation}
\label{eq:main-37}
\mathcal P_{\mathrm{nl}}^{(n)} = \sum_L\int_{\mathcal I_{L,c}} \left[ \overline{\hat G_{c,ij}}\, \hat{\mathcal S}_{ij}^{(n,m_c)} + \overline{\hat p_c}\, \hat{\mathcal D}^{(n,m_c)} \right]dx_2.
\end{equation}
We form the first term by contracting the nonlinear stress source with the complex-conjugate critical displacement gradient, and the second by multiplying the nonlinear incompressibility source by the complex-conjugate critical pressure. The hatted sources in Eq. \eqref{eq:main-37} denote the resonant coefficient fields after the associated amplitude factor has been extracted. Here \(L\) denotes the film and substrate, and \(\mathcal I_{L,c}\) is the corresponding depth interval in the critical current configuration.

At third order, the resonant forcing consists of three physically distinct contributions: slow-time inertia, loading detuning, and the cubic nonlinear interaction of the first- and second-order fields. Their projections are
\begin{equation}
\label{eq:main-projection-components}
\mathcal P_{\mathrm{in}}=\mathcal M A_{\tau\tau}, \qquad \mathcal P_{\Delta}=-\gamma_0\Delta_2A, \qquad \mathcal P_{\mathrm{nl}}=\gamma_1|A|^2A.
\end{equation}
Here the positive modal mass is \(\mathcal M=\sum_L\int_{\mathcal I_{L,c}}\rho_L\,\overline{\hat u_{c,i}}\hat u_{c,i}\,dx_2>0\).

The loading coefficient must account for how the complete linear problem changes when \(\lambda\) is varied. This variation changes the instantaneous material response and the current geometry: the film thickness \(h=H/\lambda\) and the relation \(X_2=\lambda x_2\) both depend on \(\lambda\). To compare neighbouring base states on the same domain, we first map them to the fixed critical current configuration and then differentiate the complete mixed linear sesquilinear form,
\begin{equation}
\label{eq:main-38}
\gamma_0 = \left. \frac{\partial}{\partial\lambda} \mathfrak b_\lambda \!\left[ (\hat{\mathbf u}_c,\hat p_c), (\hat{\mathbf u}_c,\hat p_c) \right] \right|_{\lambda=\lambda_c}.
\end{equation}

The explicit pull-back derivative is given in Appendix~\ref{loading-derivative-and-neo-hookean-cubic-reduction}. The cubic nonlinear projection is
\begin{equation}
\label{eq:main-39}
\gamma_1 = \sum_L\int_{\mathcal I_{L,c}} \left[ \overline{\hat G_{c,ij}}\, \hat{\mathcal S}_{ij}^{(3,m_c)} + \overline{\hat p_c}\, \hat{\mathcal D}^{(3,m_c)} \right]dx_2.
\end{equation}

The third-order Fredholm condition requires the total resonant projection to vanish,
\begin{equation}
\label{eq:main-unnormalized-cubic-amplitude}
\mathcal M A_{\tau\tau} - \gamma_0\Delta_2A + \gamma_1|A|^2A =0.
\end{equation}
For a given system \(\boldsymbol\xi\), define \(c_0(\boldsymbol\xi)=\gamma_0(\boldsymbol\xi)/\mathcal M(\boldsymbol\xi)\) and \(c_1(\boldsymbol\xi)=\gamma_1(\boldsymbol\xi)/\mathcal M(\boldsymbol\xi)\). The explicit dependence on \(\boldsymbol\xi\) is suppressed below unless needed to distinguish neighbouring systems. The cubic amplitude equation is then
\begin{equation}
\label{eq:main-40}
A_{\tau\tau} - c_0\Delta_2A + c_1|A|^2A =0.
\end{equation}

For a steady bifurcation branch, introduce the real physical modal-amplitude measure \(a=\varepsilon|A|\), with the critical-mode normalization fixed. Returning to the physical detuning, Eq. \eqref{eq:main-40} gives
\begin{equation}
\label{eq:main-41}
-c_0\Delta a+c_1a^3=0, \qquad \Delta=\frac{c_1}{c_0}a^2 \quad(a\neq0).
\end{equation}
Equation \eqref{eq:main-41} shows that \(\left.d\Delta/d(a^2)\right|_{a=0}=c_1/c_0\). For the simple stability crossing considered here, increasing compression through \(\lambda_c\) gives \(\gamma_0>0\), and hence \(c_0>0\). The coefficient \(c_1\) therefore controls the initial slope of the branch in the \((a^2,\Delta)\) plane. A positive \(c_1\) gives an initially supercritical branch, while a negative \(c_1\) gives an initially subcritical one. The corresponding cubic scaling is \(a\sim|\Delta|^{1/2}\). As \(c_1\to0\), the cubic balance becomes degenerate and the next resonant order must be retained.

\subsection{Cubic degeneracy and quintic branch selection}\label{cubic-degeneracy-quintic-branch-selection}

Within the simple isolated critical branch considered here, define the cubic-degeneracy locus by \(\Sigma_3:=\{\boldsymbol\xi:c_1(\boldsymbol\xi)=0\}\). A point \(\boldsymbol\xi^\star\in\Sigma_3\) is a simple cubic-degenerate system if \(c_1^\star:=c_1(\boldsymbol\xi^\star)=0\) and \(\nabla_{\boldsymbol\xi}c_1(\boldsymbol\xi^\star)\neq\mathbf0\). The non-zero gradient ensures that \(c_1\) changes linearly away from zero under a small parameter perturbation, so \(c_1=0\) defines a regular local transition surface separating systems with \(c_1>0\) from those with \(c_1<0\).
A superscript \(\star\) denotes evaluation at this degenerate system. For example, \(\lambda_c^\star=\lambda_c(\boldsymbol\xi^\star)\), \(m_c^\star=m_c(\boldsymbol\xi^\star)\), and \(\hat{\boldsymbol\eta}_c^\star=\hat{\boldsymbol\eta}_c(\boldsymbol\xi^\star)\). The subscript \(c\) continues to denote the linear critical state of the system under consideration. The fifth-order reduction below is evaluated at this exact degenerate system, with its material and structural parameters held fixed.

At \(\boldsymbol\xi=\boldsymbol\xi^\star\), the cubic solvability coefficient vanishes, so the third-order amplitude equation no longer determines a non-trivial local branch. The next resonant contribution occurs at fifth order. The appropriate distinguished scaling is
\begin{equation}
\label{eq:main-42}
\Delta = \lambda_c^\star-\lambda = \varepsilon^4\Delta_4, \qquad T=\varepsilon^2t.
\end{equation}
Under this scaling, slow inertia, loading detuning, and the quintic resonant forcing all enter at \(O(\varepsilon^5)\). At \(O(\varepsilon^3)\), the \(m_c^\star\)-harmonic problem still contains a generally non-zero cubic forcing. Its Fredholm projection vanishes because \(c_1^\star=0\), which guarantees solvability but does not make the forcing itself zero. A particular third-order resonant field must therefore still be computed. Thus \(c_1^\star=0\) removes the cubic solvability term but not the third-order resonant field, which is still required in the fifth-order calculation.

Because the critical operator has a one-dimensional null space, the resonant third-order field is non-unique and may be written \(\hat{\boldsymbol\eta}^{(3,m_c^\star)}=\hat{\boldsymbol\eta}^{(3,m_c^\star)}_p+\vartheta\hat{\boldsymbol\eta}_c^\star\). At \(c_1^\star=0\), a gauge fixes the remaining null-space component. We choose
\begin{equation}
\label{eq:main-43}
\sum_L\int_{\mathcal I_{L,c}^\star} \rho_L\, \overline{\hat{\mathbf u}}_c^\star \cdot \hat{\mathbf u}^{(3,m_c^\star)}\,dx_2 =0.
\end{equation}
This fixes the amplitude convention; the corresponding invariance of \(c_2^\star\) is shown in Appendix~\ref{fifth-order-solvability-at-cubic-degeneracy}. This gauge removes an arbitrary addition of the critical mode at third order; it does not change the normalization of the leading eigenfield. The harmonic chain required to construct the fifth-order critical forcing is
\begin{equation}
\label{eq:main-harmonic-sets-at-degeneracy}
\mathcal Q_3=\{\pm m_c^\star,\pm3m_c^\star\}, \qquad \mathcal Q_4=\{0,\pm2m_c^\star,\pm4m_c^\star\}, \qquad \mathcal Q_5=\{\pm m_c^\star,\pm3m_c^\star,\pm5m_c^\star\}.
\end{equation}

Together with the first- and second-order fields, the third-order \(m_c^\star\) and \(3m_c^\star\) fields and the fourth-order \(0\), \(2m_c^\star\), and \(4m_c^\star\) fields determine the fifth-order \(m_c^\star\)-forcing. Its Fredholm projection defines
\begin{equation}
\label{eq:main-44}
\gamma_2^\star = \sum_L\int_{\mathcal I_{L,c}^\star} \left[ \overline{\hat G_{c,ij}^\star}\, \hat{\mathcal S}_{ij}^{(5,m_c^\star)} + \overline{\hat p_c^\star}\, \hat{\mathcal D}^{(5,m_c^\star)} \right]dx_2, \qquad c_2^\star=\frac{\gamma_2^\star}{\mathcal M^\star}.
\end{equation}
The exact cubic-degenerate amplitude equation is therefore
\begin{equation}
\label{eq:main-45}
A_{TT} - c_0^\star\Delta_4A + c_2^\star|A|^4A =0.
\end{equation}
Returning to the physical modal amplitude \(a=\varepsilon|A|\), the steady branch satisfies
\begin{equation}
\label{eq:main-46}
-c_0^\star\Delta a+c_2^\star a^5=0, \qquad \Delta=\frac{c_2^\star}{c_0^\star}a^4 \quad(a\neq0).
\end{equation}
Thus \(c_2^\star\) supplies the first non-vanishing nonlinear branch information at \(c_1^\star=0\), replacing the generic square-root scaling by \(a\sim|\Delta|^{1/4}\). Since \(c_0^\star>0\), \(c_2^\star>0\) sends the exact degenerate branch toward the post-critical (supercritical) side \(\Delta>0\), whereas \(c_2^\star<0\) sends it toward the pre-critical (subcritical) side \(\Delta<0\). The fifth-order classification requires \(c_2^\star\neq0\). If \(c_1^\star=c_2^\star=0\), the retained nonlinear balance is also degenerate and higher resonant orders are required.

\begin{figure}[htbp]
\centering
\includegraphics[width=0.8\textwidth]{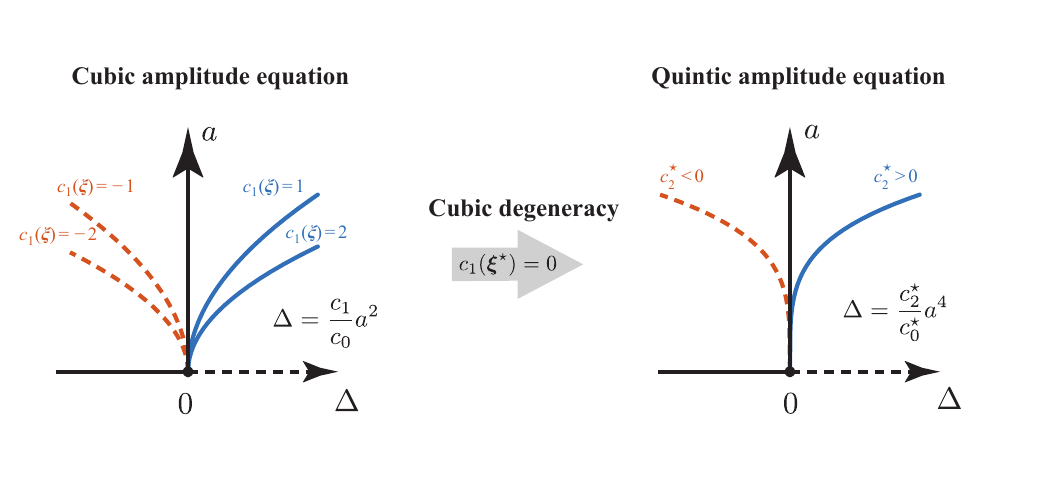}
\caption{Schematic local relations between modal amplitude and compression detuning for the cubic amplitude equation and the fifth-order amplitude equation at cubic degeneracy. Here \(\Delta=\lambda_c-\lambda\) measures the loading departure from the critical state and \(a\) is the modal amplitude. Left: for \(c_1\neq0\), the cubic amplitude equation gives \(\Delta=(c_1/c_0)a^2\); \(c_1>0\) and \(c_1<0\) correspond to supercritical and subcritical branches, respectively. Right: at the cubic-degenerate state \(\boldsymbol\xi^\star\), where \(c_1^\star=0\), the fifth-order amplitude equation gives \(\Delta=(c_2^\star/c_0^\star)a^4\); the sign of \(c_2^\star\) determines whether the degenerate branch is supercritical or subcritical.}
\label{fig:figure4}
\end{figure}

Figure~\ref{fig:figure4} compares the steady cubic amplitude equation with the fifth-order amplitude equation required at cubic degeneracy. For \(c_1\neq0\), the cubic equation gives \(\Delta\propto a^2\), and the sign of \(c_1\) determines whether the branch is supercritical or subcritical. At \(c_1^\star=0\), the cubic term vanishes and the fifth-order equation gives \(\Delta\propto a^4\); the sign of \(c_2^\star\) then determines whether the degenerate branch is supercritical or subcritical.

\FloatBarrier
\section{Results for a graded neo-Hookean film--substrate system}\label{results-for-a-graded-neo-hookean-filmsubstrate-system}

We apply the higher-order Stroh framework to an incompressible neo-Hookean film bonded to an exponentially graded neo-Hookean substrate. The analysis examines how substrate gradation affects the nonlinear bifurcation beyond the linear wrinkling threshold. After specifying the graded neo-Hookean model, we validate the linear and cubic calculations against established results. We then examine how the gradation parameters shift the cubic transition \(c_1=0\). Finally, the analysis is continued to fifth order on the resulting cubic-degeneracy surface in order to determine the branch direction when the cubic coefficient vanishes.

\subsection{Neo-Hookean system and validation}\label{neo-hookean-system-and-validation}

Both the film and substrate are taken to be incompressible neo-Hookean materials, with strain-energy density \(W=\mu[\operatorname{tr}(\mathbf F^{\mathrm T}\mathbf F)-3]/2\), where \(\mathbf F\) is the three-dimensional deformation gradient (\(F_{33}=1\) in plane strain). The film is homogeneous with shear modulus \(\mu_f\). In the substrate, the shear modulus varies continuously with the reference depth \(X_2\) according to
\begin{equation}
\label{eq:main-51}
\mu(X_2) = \mu_s \left[ \beta+(1-\beta)e^{\alpha X_2} \right], \qquad X_2\le0,
\end{equation}
where \(\mu_s=\mu(0)\) is the substrate shear modulus at the film--substrate interface. The parameter \(\alpha\ge0\) controls the rate of variation with depth, while the far-field modulus ratio \(\beta=\mu_\infty/\mu_s\) lies in \(0<\beta\le1\), with \(\mu_\infty\) denoting the far-field substrate shear modulus. Thus, \(\beta<1\) corresponds to a substrate that becomes softer away from the interface. Either \(\alpha=0\) or \(\beta=1\) removes the depth dependence and recovers a homogeneous substrate.

The grading law in Eq. \eqref{eq:main-51} is prescribed in the reference configuration. Under the homogeneous plane-strain base deformation, \(X_2=\lambda x_2\), so the corresponding modulus profile in the current configuration is
\begin{equation}
\label{eq:current-grading-profile}
\mu(x_2)
=
\mu_s
\left[
\beta+(1-\beta)e^{\alpha\lambda x_2}
\right].
\end{equation}
We use \(r=\mu_s/\mu_f\) to measure the substrate-to-film shear modulus ratio and \(\alpha H\) as the dimensionless decay rate. The graded film--substrate system is therefore characterized by
\begin{equation}
\label{eq:grading-parameter-vector}
\boldsymbol\xi=(\alpha H,\beta,r).
\end{equation}

For the neo-Hookean strain-energy function, \(W\) is quadratic in \(\mathbf F\). Consequently,
\begin{equation}
\label{eq:neo-hookean-higher-moduli}
\mathcal A^s=0,
\qquad
s\ge2.
\end{equation}
The higher constitutive tensors therefore do not contribute to the nonlinear source terms. The nonlinear contributions at second and higher orders arise instead from the incompressibility constraint and the associated pressure fields generated by the lower-order solutions.

At cubic order, this structure leads to a substantial simplification. The resonant third-order projection can be expressed entirely in terms of the critical field and the second-order mean and second-harmonic corrections. Let \(G_{ij}=\hat G_{c,ij}\), \(G_{0,ij}=\hat G_{ij}^{(2,0)}\), and \(G_{2,ij}=\hat G_{ij}^{(2,2m_c)}\), and let \(\pi=\hat p_c\), \(\pi_0=\hat p^{(2,0)}\), and \(\pi_2=\hat p^{(2,2m_c)}\). The general cubic projection in Eq. \eqref{eq:main-39} then reduces, for the present two-dimensional incompressible neo-Hookean system, to
\begin{equation}
\label{eq:main-52}
\gamma_1 = \sum_L\int_{\mathcal I_{L,c}} \Big\{ \overline G_{ij} \left( \pi G_{0,ji} + \overline\pi\,G_{2,ji} + \pi_0G_{ji} + \pi_2\overline G_{ji} \right)  + \overline\pi \left( G_{ij}G_{0,ji} + \overline G_{ij}G_{2,ji} \right) \Big\}\,dx_2 .
\end{equation}
Equation \eqref{eq:main-52} makes explicit which fields determine the cubic coefficient. Once the critical mode and the second-order \(0\) and \(2m_c\) fields have been computed, \(\gamma_1\) follows by integrating Eq. \eqref{eq:main-52} over the film and substrate, and the cubic coefficient is obtained from \(c_1=\gamma_1/\mathcal M\). The derivation of Eq. \eqref{eq:main-52} from the general projection in Eq. \eqref{eq:main-39} is given in Appendix~\ref{loading-derivative-and-neo-hookean-cubic-reduction}.

The critical and second-order fields are obtained numerically from the corresponding linear and forced Stroh problems. A fixed normalization of the leading critical mode is used throughout. A positive rescaling of the critical mode changes the numerical magnitude of \(c_1\), but not its sign or the location of the zero \(c_1=0\). At a particular cubic-degenerate system \(\boldsymbol\xi^\star\in\Sigma_3\), defined by \(c_1^\star=c_1(\boldsymbol\xi^\star)=0\), the second-order fields are no longer sufficient to determine the leading nonlinear branch. The quintic coefficient \(c_2^\star\) additionally requires the gauge-fixed third-order \(m_c^\star\) field, the third-order \(3m_c^\star\) field, and the fourth-order \(0\), \(2m_c^\star\), and \(4m_c^\star\) fields, as described in Section 4.3. These fields are then used to construct the fifth-order resonant source entering Eq. \eqref{eq:main-44}.

Before examining the effect of gradation on \(c_1\) and \(c_2^\star\), we validate the numerical implementation at two independent levels. The linear calculation is tested against the graded-substrate results of \citet{liu2024}, whereas the cubic calculation is tested in the homogeneous-substrate limit against the classical result of \citet{caifu1999}. These two comparisons examine different parts of the formulation: the former tests the variable-coefficient linear bifurcation problem, while the latter tests the nonlinear sequence from the second-order forced fields to the third-order Fredholm projection.

\begin{figure}[!htbp]
\centering
\includegraphics[width=0.98\textwidth]{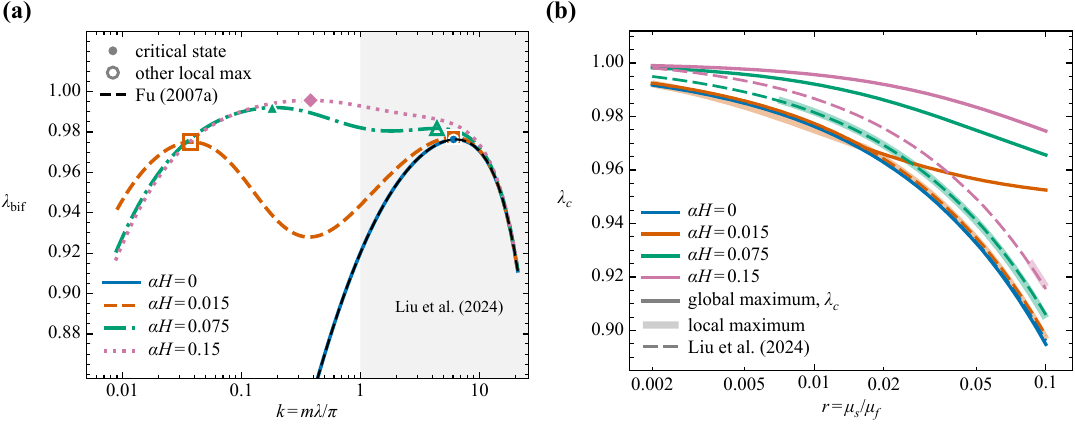}
\caption{Linear bifurcation benchmarks for the graded film--substrate system at \(\beta=0.01\). (a) Bifurcation stretch \(\lambda_{\mathrm{bif}}\) versus the benchmark wavenumber coordinate \(k=m\lambda/\pi\) for \(\alpha H=0,0.015,0.075,0.15\). The shaded wavenumber interval indicates the root-search range used in \citet{liu2024}. The local maxima of each dispersion curve are identified, and the larger maximum gives the selected critical state. (b) Selected critical stretch \(\lambda_c=\max_m\lambda_{\mathrm{bif}}\) versus the substrate-to-film modulus ratio \(r=\mu_s/\mu_f\) for the same grading rates. Solid curves show the present critical stretches, faint curves retain the competing local-max branches, and dashed curves denote the reference results of \citet{liu2024}.}
\label{fig:figure5}
\end{figure}

Figure~\ref{fig:figure5}(a) shows the bifurcation stretch \(\lambda_{\mathrm{bif}}\) as a function of the wavenumber \(k\) for several decay rates \(\alpha H=0,0.015,0.075,0.15\) at \(\beta=0.01\). When the dispersion curve contains a single maximum, the critical state is obtained directly from that maximum. For some parameter combinations, however, two local maxima appear within the search range. In this case, following only one branch can identify a local maximum that is not the true critical state. Since compression increases as \(\lambda\) decreases from unity, the instability that occurs first is associated with the largest value of \(\lambda_{\mathrm{bif}}\) over the entire search range. We therefore retain all candidate maxima and select the global maximum to determine the critical stretch \(\lambda_c\) and the corresponding critical wavenumber \(m_c\).

Figure~\ref{fig:figure5}(b) compares the critical stretches obtained from this global search with the results of \citet{liu2024}. The two results agree when the maximum identified within the reference search interval is also the global maximum. In parameter ranges where another branch attains a larger value of \(\lambda_{\mathrm{bif}}\), the reference result follows a local maximum rather than the first instability of the full dispersion relation. The present search captures this branch switching by comparing all local maxima within the prescribed wavenumber range and selecting the largest one. The resulting \(\lambda_c(r)\) therefore represents the actual onset of instability over the full search range considered.

In the homogeneous-substrate limit \((\alpha=0)\), we denote by \(r^\star\) the shear modulus ratio satisfying \(c_1(r^\star)=0\). Since \(c_0>0\), Eq. \eqref{eq:main-41} shows that \(r^\star\) separates the two local post-bifurcation responses: \(c_1>0\) corresponds to a supercritical branch, whereas \(c_1<0\) corresponds to a subcritical branch.

\begin{figure}[!htbp]
\centering
\includegraphics[width=0.6\textwidth]{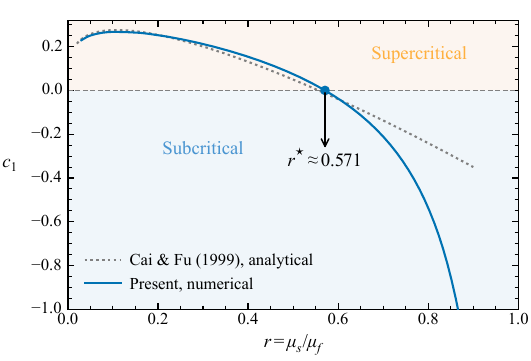}
\caption{Cubic benchmark for the homogeneous-substrate limit \((\alpha=0)\). The computed cubic coefficient \(c_1\) is plotted against the substrate-to-film stiffness ratio \(r=\mu_s/\mu_f\) and compared with the analytical result of \citet{caifu1999}. The cubic transition occurs at \(r=r^\star\simeq0.571\), where \(c_1=0\). For \(r<r^\star\), \(c_1>0\) and the primary branch is supercritical; for \(r>r^\star\), \(c_1<0\) and the branch is subcritical. The transition point \(r^\star\) is marked in the figure.}
\label{fig:figure6}
\end{figure}

Figure~\ref{fig:figure6} considers the homogeneous-substrate limit obtained by setting \(\alpha=0\). In this case, Eq. \eqref{eq:main-51} reduces to \(\mu(X_2)=\mu_s\), so the substrate modulus is independent of depth and \(\beta\) no longer affects the modulus profile. The present calculation gives \(r^\star\simeq0.571\), in agreement with \citet{caifu1999}. Near the transition, the present numerical result agrees closely with the analytical approximation. As \(r\) increases further into the subcritical regime, the two curves gradually separate. A similar difference between the numerical and asymptotic results was reported by \citet{caifu1999} at larger \(r\). The analytical expression is based on an asymptotic approximation and becomes less accurate as the system moves away from the corresponding asymptotic regime, while the present calculation follows the same numerical trend. The comparison therefore reproduces both the transition at \(r^\star\) and the change from supercritical to subcritical response across this transition.

\subsection{Gradation-controlled cubic transition}\label{gradation-controlled-cubic-transition}
We extend the transition ratio \(r^\star\) defined above to the graded substrate. For each prescribed pair \((\alpha H,\beta)\), the transition value \(r^\star=r^\star(\alpha H,\beta)\) is determined by
\begin{equation}
\label{eq:cubic-transition-definition}
c_1(\alpha H,\beta,r^\star)=0.
\end{equation}
The corresponding cubic-degenerate system is therefore \(\boldsymbol\xi^\star=(\alpha H,\beta,r^\star)\). As in the homogeneous case, \(c_1>0\) and \(c_1<0\) correspond to supercritical and subcritical branches, respectively.

\begin{figure}[htbp]
\centering
\includegraphics[width=0.72\textwidth]{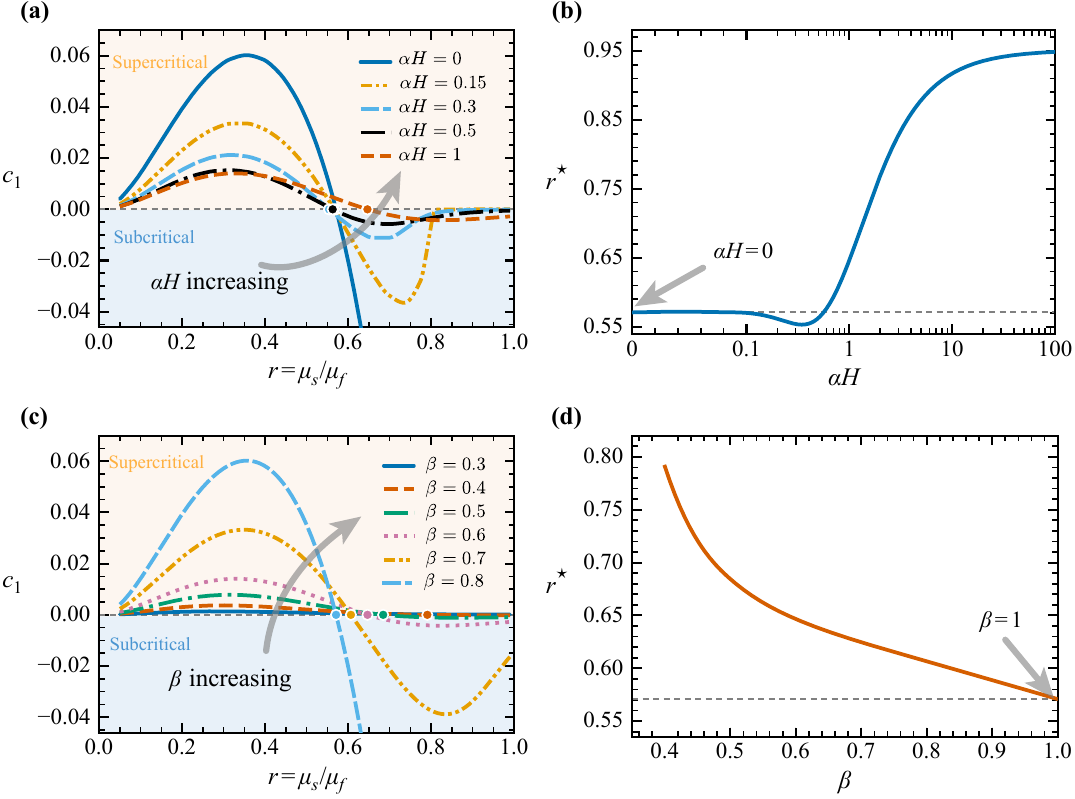}
\caption{Effect of substrate gradation on the cubic transition. (a) \(c_1(r)\) at fixed \(\beta=0.6\) for \(\alpha H=0,\,0.15,\,0.3,\,0.5,\) and \(1\). (b) Corresponding transition ratio \(r^\star\) as a function of \(\alpha H\); the homogeneous-substrate value at \(\alpha H=0\) is indicated separately. (c) \(c_1(r)\) at fixed \(\alpha H=1\) for \(\beta=0.3,\,0.4,\,0.5,\,0.6,\,0.7,\) and \(0.8\). (d) Corresponding transition ratio \(r^\star(\beta)\), which approaches the homogeneous-substrate value as \(\beta\to1\). In (a,c), the horizontal zero line separates the supercritical \((c_1>0)\) and subcritical \((c_1<0)\) regions.}
\label{fig:figure7}
\end{figure}

At fixed \(\beta=0.6\), Fig.~\ref{fig:figure7}(a) shows that \(c_1\) first increases with \(r\), reaches a positive maximum, and then decreases through zero at \(r=r^\star\). Increasing \(\alpha H\) reduces the positive peak and shifts the zero crossing. The corresponding variation of \(r^\star\) is summarized in Fig.~\ref{fig:figure7}(b). Starting from the homogeneous value \(r^\star\simeq0.571\) at \(\alpha H=0\), \(r^\star\) first decreases slightly, reaching a minimum in the weak-grading regime, and then increases rapidly with \(\alpha H\), approaching about \(0.95\) for large \(\alpha H\). The cubic transition therefore depends non-monotonically on the decay rate \(\alpha H\).

Figures~\ref{fig:figure7}(c,d) show the complementary effect of the far-field modulus ratio at fixed \(\alpha H=1\). As \(\beta\) increases, the positive part of \(c_1(r)\) becomes larger and the zero crossing shifts toward smaller \(r\). Consequently, \(r^\star\) decreases monotonically with \(\beta\), from approximately \(0.79\) at \(\beta=0.4\) toward \(r^\star\simeq0.571\) as \(\beta\to1\). The latter limit corresponds to a homogeneous substrate, because \(\beta=1\) removes the depth dependence of the modulus profile.

Referring to Eq. \eqref{eq:main-52}, the gradation parameters \(\alpha H\) and \(\beta\) do not control \(c_1\) directly. Instead, they modify the stiffness distribution through the substrate depth, which changes the critical mode and the second-order mean field and second-harmonic correction entering the cubic projection. These changes alter \(\gamma_1\), and hence \(c_1=\gamma_1/\mathcal M\). Since \(\mathcal M>0\), the condition \(c_1=0\) means that the integrated nonlinear contributions cancel overall. The results in Fig.~\ref{fig:figure7} therefore show that the cubic transition depends on the complete depth-dependent stiffness profile of the substrate as well as the interfacial shear modulus ratio \(r\).

\FloatBarrier
\subsection{Cubic-degeneracy surface and quintic branch selection}\label{cubic-degeneracy-surface-and-quintic-branch-selection}

\begin{figure}[!htbp]
\centering
\includegraphics[width=\textwidth]{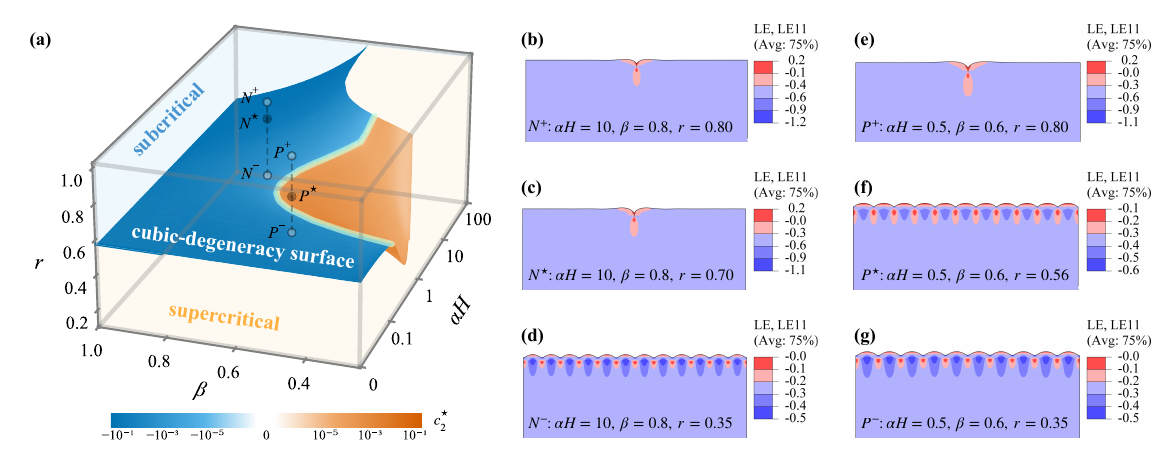}
\caption{Cubic-degeneracy surface, quintic branch selection, and representative finite-element deformation fields. (a) Cubic-degeneracy surface \(c_1(\alpha H,\beta,r^\star)=0\) in the parameter space \((\alpha H,\beta,r)\). The light-orange and light-blue regions on the two sides of the surface correspond to the supercritical \((c_1>0)\) and subcritical \((c_1<0)\) regimes, respectively. On the degeneracy surface, the colour represents \(c_2^\star\): orange denotes \(c_2^\star>0\) and blue denotes \(c_2^\star<0\). Their boundary \(c_2^\star=0\) defines the quintic-degeneracy curve. The marked points \(N^\star\) and \(P^\star\) represent negative- and positive-\(c_2^\star\) cubic-degenerate states, respectively. (b--d) Longitudinal logarithmic strain fields for the \(N\) family with \(\alpha H=10\), \(\beta=0.8\), and \(r=0.80\), \(0.70\), and \(0.35\), corresponding to \(N^+\), \(N^\star\), and \(N^-\), respectively. (e--g) Corresponding fields for the \(P\) family with \(\alpha H=0.5\), \(\beta=0.6\), and \(r=0.80\), \(0.56\), and \(0.35\), corresponding to \(P^+\), \(P^\star\), and \(P^-\), respectively.}
\label{fig:figure8}
\end{figure}

We examine the fifth-order response on the cubic-degeneracy surface \(\Sigma_3\) introduced in Section 4.3. Every point on this surface satisfies \(c_1=0\), so the cubic amplitude equation no longer determines whether the local branch is supercritical or subcritical. We therefore evaluate the quintic coefficient \(c_2^\star\) over the entire surface. Figure~\ref{fig:figure8}(a) shows the resulting surface \(r^\star(\alpha H,\beta)\) in the three-dimensional parameter space \((\alpha H,\beta,r)\). The light-orange and light-blue regions on the two sides of the surface correspond to \(c_1>0\) and \(c_1<0\), respectively. On the surface itself, the colour represents \(c_2^\star\): the orange part has \(c_2^\star>0\) and is supercritical at cubic degeneracy, whereas the blue part has \(c_2^\star<0\) and is subcritical. Substrate gradation therefore affects the location of the cubic transition and the branch direction selected after the cubic term vanishes.

Within \(\Sigma_3\), the boundary between the \(c_2^\star>0\) and \(c_2^\star<0\) regions satisfies \(c_2^\star=0\), defining a quintic-degeneracy curve. Along this curve, both the cubic and quintic nonlinear coefficients vanish, so the fifth-order theory is also degenerate and the branch direction requires a seventh-order, or higher-order, analysis. To illustrate the two branch selections on either side of this curve, we choose two representative cubic-degenerate states. The point \(N^\star\), approximately located at \((\alpha H,\beta,r^\star)=(10,0.8,0.70)\), lies in the \(c_2^\star<0\) region and is subcritical, whereas \(P^\star\), approximately located at \((0.5,0.6,0.56)\), lies in the \(c_2^\star>0\) region and is supercritical. Both satisfy \(c_1=0\), so their different branch directions cannot be distinguished by the cubic theory and only become apparent after \(c_2^\star\) is evaluated.

For each of these two parameter families, \(r\) is further varied across the cubic transition while \(\alpha H\) and \(\beta\) are kept fixed. In the \(N\) family, \(N^+\) has \(r=0.80>r^\star\) and lies in the \(c_1<0\) subcritical region, \(N^\star\) lies on \(c_1=0\) with \(c_2^\star<0\), and \(N^-\) has \(r=0.35<r^\star\) and lies in the \(c_1>0\) supercritical region. Similarly, \(P^+\) is subcritical, \(P^\star\) lies on \(c_1=0\) with \(c_2^\star>0\), and \(P^-\) is supercritical. The superscripts \(+\) and \(-\) indicate whether \(r\) is above or below \(r^\star\), rather than the sign of either nonlinear coefficient.

The finite-element model is implemented in Abaqus and uses shared nodes at the film--substrate interface and four-node hybrid plane-strain elements (CPE4H), with mesh refinement in the film and near the interface. The depth-dependent substrate modulus is implemented through a user-defined incompressible hyperelastic material. Horizontal compression is imposed through a reference point with periodic lateral boundary conditions; the bottom boundary is restrained vertically, and centreline and mirror-symmetry constraints are applied. Nonlinear post-buckling calculations use a static analysis step with damping and self-contact. Initial geometric imperfections are introduced using linear buckling modes for wrinkling and a symmetric local notch for creasing.

Finite-element simulations are performed for the six representative states marked in Fig.~\ref{fig:figure8}(a), and the corresponding longitudinal logarithmic strain fields are shown in Figs.~\ref{fig:figure8}(b--g). Within each family, \(\alpha H\) and \(\beta\) are fixed while \(r\) is varied from the subcritical side, through the cubic-degenerate state, to the supercritical side. The subcritical states \(N^+\) and \(P^+\) show strong localization around a dominant surface indentation, and the cubic-degenerate state \(N^\star\), for which \(c_2^\star<0\), exhibits a similar tendency. By contrast, the supercritical states \(N^-\) and \(P^-\) retain a distributed periodic wrinkling pattern, while the cubic-degenerate state \(P^\star\), for which \(c_2^\star>0\), likewise develops a periodic multi-wrinkle morphology.

These finite-element results are consistent with the local branch classifications obtained from \(c_1\) and \(c_2^\star\). Over the deformation range considered, the subcritical cases exhibit a stronger tendency toward localization, whereas the supercritical cases retain the periodic wrinkling mode and evolve through a more distributed increase in wrinkle amplitude. The simulations therefore provide a finite-amplitude illustration of the distinct post-bifurcation responses associated with the two branch types.

\section{Discussion and conclusions}\label{sec:discussion}\label{sec:conclusions}

This paper develops a higher-order Stroh framework for weakly nonlinear post-bifurcation analysis of finitely deformed hyperelastic solids. For each nonzero harmonic, the same base-state Stroh operator is retained across perturbation orders, while nonlinear source terms are assembled from lower-order fields. The amplitude equations are obtained by imposing Fredholm solvability conditions at the resonant orders. At cubic degeneracy, the third-order resonant field generally remains nonzero even though the cubic amplitude coefficient vanishes, and it contributes to the fifth-order source terms.

For an incompressible neo-Hookean film bonded to an exponentially graded substrate, gradation shifts the supercritical--subcritical transition by modifying the critical mode and the lower-order fields entering the cubic solvability condition. The interfacial modulus ratio alone is therefore insufficient to determine the transition. In the homogeneous-substrate limit, the computed transition agrees with the classical result of \citet{caifu1999}. The states satisfying \(c_1=0\) form a cubic-degeneracy surface in the parameter space of modulus ratio and substrate gradation. On this surface, \(c_2^\star<0\) gives a locally subcritical branch, whereas \(c_2^\star>0\) gives a locally supercritical branch. Finite-element simulations at two selected cubic-degenerate states show that the subcritical state develops a surface crease, whereas the supercritical state retains a periodic wrinkling pattern (Fig.~\ref{fig:figure8}). The two regions meet along a quintic-degeneracy curve where \(c_1=c_2^\star=0\). On this curve, the fifth-order theory no longer determines the branch direction, and the expansion must be continued to seventh order or beyond.

The framework provides a systematic way to compute higher-order corrections and determine local branch directions near a simple, isolated critical mode. It can be adapted to other instability problems, such as buckling in multilayer solids and growth-induced instabilities in soft tissues. Beyond bifurcation analysis, the recursive Stroh formulation also offers a route to higher-order indentation analysis, with the appropriate spatial representation and contact boundary conditions.

\section*{Acknowledgements}
This work was supported by the National Natural Science Foundation of China (Grant Nos. 12402197, 12602263, 125B2042, and 12672131). The work was also partially supported by the specialized research projects of Huanjiang Laboratory, Zhuji, Zhejiang Province, China, and the 111 Project, China (No. B21034).

\section*{CRediT authorship contribution statement}
\textbf{Shiyuan Deng:} Methodology, Software, Validation, Investigation, Writing -- original draft.
\textbf{Han Ye:} Validation, Investigation.
\textbf{Yafei Wang:} Writing -- review \& editing.
\textbf{Lizichen Chen:} Writing -- review \& editing.
\textbf{Alain Goriely:} Writing -- review \& editing.
\textbf{Yibin Fu:} Writing -- review \& editing.
\textbf{Weiqiu Chen:} Writing -- review \& editing.
\textbf{Chaofeng L\"u:} Writing -- review \& editing.
\textbf{Michel Destrade:} Conceptualization, Methodology, Writing -- review \& editing.
\textbf{Kecheng Li:} Methodology, Writing -- review \& editing.
\textbf{Yangkun Du:} Conceptualization, Methodology, Writing -- original draft, Writing -- review \& editing, Funding acquisition.

\section*{Declaration of competing interest}
The authors declare that they have no known competing financial interests or personal relationships that could have appeared to influence the work reported in this paper.

\appendix
\numberwithin{equation}{section}
\renewcommand{\theequation}{\Alph{section}.\arabic{equation}}
\renewcommand{\theHequation}{App.\Alph{section}.\arabic{equation}}
\renewcommand{\thesection}{\Alph{section}}
\makeatletter
\let\manuscriptsectionformat\@seccntformat
\renewcommand{\@seccntformat}[1]{%
  \ifstrequal{#1}{section}{Appendix~\csname the#1\endcsname.\quad}{\manuscriptsectionformat{#1}}}
\makeatother

\section{Higher-order nonlinear sources}\label{higher-order-nonlinear-sources}

The recursive Stroh hierarchy requires the nonlinear stress and incompressibility sources at each perturbation order. This section gives their construction using the notation of Sections 2 and 3 of the main text. The incremental displacement gradient is \(\mathbf G=\operatorname{grad}\mathbf u\), and its perturbation coefficients satisfy \(\mathbf G^{(n)}=\operatorname{grad}\mathbf u^{(n)}\). The order-\(n\) stress is written, as in Eq. \eqref{eq:main-10} of the main text, as
\begin{equation}
\label{eq:sm-order-n-stress-decomposition}
\chi^{(n)}_{ij} = \bar{\mathcal A}^{1}_{jilk}G^{(n)}_{kl} - p^{(n)}\delta_{ij} + \mathcal S^{(n)}_{ij},
\end{equation}
where \(\bar{\mathcal A}^{1}_{jilk}=\mathcal A^{1}_{jilk}+\bar p\,\delta_{jk}\delta_{il}\). Thus the unknown order-\(n\) fields appear only in the linear part. The nonlinear source \(\mathcal S^{(n)}_{ij}\) is assembled from lower-order fields.

To avoid confusion between perturbation order and matrix multiplication, the notation \(\mathbf G^{(n)}\) is used only for the order-\(n\) perturbation gradient. Products of gradient tensors are written explicitly. In components,
\begin{equation}
\label{eq:sm-S1}
\left[ \mathbf G^{(a_1)} \mathbf G^{(a_2)} \cdots \mathbf G^{(a_s)} \right]_{ji} = G^{(a_1)}_{jk_1} G^{(a_2)}_{k_1k_2} \cdots G^{(a_s)}_{k_{s-1}i}.
\end{equation}

Extending the definition in Eq. \eqref{eq:main-4} of the main text, the higher instantaneous tensors are written for \(s\ge1\) as
\begin{equation}
\label{eq:sm-S2}
\mathcal A^{s}_{ji\,l_1k_1\cdots l_sk_s} ={} \bar J^{-1} \bar F_{jA_0} \bar F_{l_1A_1}\cdots \bar F_{l_sA_s}  \times \frac{\partial^{s+1}W} {\partial F_{iA_0} \partial F_{k_1A_1}\cdots \partial F_{k_sA_s}} \bigg|_{\bar{\mathbf F}} .
\end{equation}
For \(s=1,2,3\), Eq. \eqref{eq:sm-S2} gives the tensors \(\mathcal A^1\), \(\mathcal A^2\), and \(\mathcal A^3\) already defined in the main text. A calculation through perturbation order \(n\) requires constitutive derivatives of \(W\) through order \(n+1\). A fifth-order calculation involves \(\mathcal A^1,\ldots,\mathcal A^5\).

The pressure contribution follows from
\begin{equation}
\label{eq:sm-S3}
(\mathbf I+\mathbf G)^{-\mathrm{T}} = \mathbf I - \mathbf G^{\mathrm{T}} + (\mathbf G\mathbf G)^{\mathrm{T}} - (\mathbf G\mathbf G\mathbf G)^{\mathrm{T}} +\cdots .
\end{equation}
After the linear terms \(\bar{\mathcal A}^{1}_{jilk}G_{kl}-\tilde p\,\delta_{ij}\) have been separated, the nonlinear stress source is
\begin{equation}
\label{eq:sm-S4}
\mathcal S_{ij} ={} \sum_{s\ge2} \frac{1}{s!} \mathcal A^{s}_{ji\,l_1k_1\cdots l_sk_s} G_{k_1l_1}\cdots G_{k_sl_s}  + \bar p \sum_{s\ge2} (-1)^{s+1} \left[ \underbrace{\mathbf G\mathbf G\cdots\mathbf G}_{s\ \mathrm{factors}} \right]_{ji}  + \tilde p \sum_{s\ge1} (-1)^{s+1} \left[ \underbrace{\mathbf G\mathbf G\cdots\mathbf G}_{s\ \mathrm{factors}} \right]_{ji}.
\end{equation}
Two different notions of order appear in Eq. \eqref{eq:sm-S4}. The integer \(s\) is the number of gradient factors in a nonlinear term. The superscript \(n\) in \(\mathbf G^{(n)}\), \(p^{(n)}\), and \(\mathcal S^{(n)}\) is the perturbation order in \(\varepsilon\). These are not the same. For example, \(\mathbf G^{(2)}\mathbf G^{(3)}\) contains two gradient factors but contributes at perturbation order five. Superscripts in parentheses always denote perturbation order in this paper.

Collecting the coefficient of \(\varepsilon^n\) gives the arbitrary-order source
\begin{equation}
\label{eq:sm-S5}
\begin{aligned} \mathcal S^{(n)}_{ij} ={}& \sum_{s=2}^{n} \frac{1}{s!} \sum_{\substack{a_1,\ldots,a_s\ge1\\ a_1+\cdots+a_s=n}} \mathcal A^{s}_{ji\,l_1k_1\cdots l_sk_s} G^{(a_1)}_{k_1l_1}\cdots G^{(a_s)}_{k_sl_s} + \bar p \sum_{s=2}^{n} (-1)^{s+1} \sum_{\substack{a_1,\ldots,a_s\ge1\\ a_1+\cdots+a_s=n}} \left( \mathbf G^{(a_1)} \cdots \mathbf G^{(a_s)} \right)_{ji} \\ &+ \sum_{\ell=1}^{n-1} p^{(\ell)} \sum_{s=1}^{n-\ell} (-1)^{s+1} \sum_{\substack{a_1,\ldots,a_s\ge1\\ a_1+\cdots+a_s=n-\ell}} \left( \mathbf G^{(a_1)} \cdots \mathbf G^{(a_s)} \right)_{ji}. \end{aligned}
\end{equation}
The terms proportional to \(\bar p\) come from the base-state pressure. The terms proportional to \(p^{(\ell)}\) come from the incremental pressure \(\tilde p=\sum_{\ell\ge1}\varepsilon^\ell p^{(\ell)}\). Thus the incremental-pressure terms satisfy \(\ell+a_1+\cdots+a_s=n\). Every quantity on the right-hand side is known once all fields below order \(n\) have been determined.

The explicit stress sources through fifth order are given below.
\begin{equation}
\label{eq:sm-S6}
\mathcal S^{(2)}_{ij} ={} \frac12 \mathcal A^2_{jilk\,nm} G^{(1)}_{kl}G^{(1)}_{mn} - \bar p \left[ (\mathbf G^{(1)})^2 \right]_{ji} + p^{(1)}G^{(1)}_{ji}.
\end{equation}
\begin{equation}
\label{eq:sm-S7}
\begin{aligned}
\mathcal S^{(3)}_{ij}
={}&
\mathcal A^2_{jilk\,nm}
G^{(1)}_{kl}G^{(2)}_{mn}
+
\frac16
\mathcal A^3_{jilk\,nm\,qp}
G^{(1)}_{kl}G^{(1)}_{mn}G^{(1)}_{pq}
-
\bar p
\left[
\mathbf G^{(1)}\mathbf G^{(2)}
+
\mathbf G^{(2)}\mathbf G^{(1)}
\right]_{ji}
+
\bar p
\left[
(\mathbf G^{(1)})^3
\right]_{ji}
\\
&+
p^{(1)}G^{(2)}_{ji}
+
p^{(2)}G^{(1)}_{ji}
-
p^{(1)}
\left[
(\mathbf G^{(1)})^2
\right]_{ji}.
\end{aligned}
\end{equation}
\begin{equation}
\label{eq:sm-S8}
\begin{aligned}
\mathcal S^{(4)}_{ij}
={}&
\mathcal A^2_{jilk\,nm}
G^{(1)}_{kl}G^{(3)}_{mn}
+
\frac12
\mathcal A^2_{jilk\,nm}
G^{(2)}_{kl}G^{(2)}_{mn}
+
\frac12
\mathcal A^3_{jilk\,nm\,qp}
G^{(1)}_{kl}G^{(1)}_{mn}G^{(2)}_{pq}
\\ &+ \frac1{24}
\mathcal A^4_{ji\,l_1k_1l_2k_2l_3k_3l_4k_4}
G^{(1)}_{k_1l_1}
G^{(1)}_{k_2l_2}
G^{(1)}_{k_3l_3}
G^{(1)}_{k_4l_4}
\\
&-
\bar p
\left[
\mathbf G^{(1)}\mathbf G^{(3)}
+
(\mathbf G^{(2)})^2
+
\mathbf G^{(3)}\mathbf G^{(1)}
\right]_{ji}
+
\bar p
\left[
(\mathbf G^{(1)})^2\mathbf G^{(2)}
+
\mathbf G^{(1)}\mathbf G^{(2)}\mathbf G^{(1)}
+
\mathbf G^{(2)}(\mathbf G^{(1)})^2
\right]_{ji}
\\ &-
\bar p
\left[
(\mathbf G^{(1)})^4
\right]_{ji}
\\
&+
p^{(1)}G^{(3)}_{ji}
+
p^{(2)}G^{(2)}_{ji}
+
p^{(3)}G^{(1)}_{ji}
-
p^{(1)}
\left[
\mathbf G^{(1)}\mathbf G^{(2)}
+
\mathbf G^{(2)}\mathbf G^{(1)}
\right]_{ji}
-
p^{(2)}
\left[
(\mathbf G^{(1)})^2
\right]_{ji}
+
p^{(1)}
\left[
(\mathbf G^{(1)})^3
\right]_{ji}.
\end{aligned}
\end{equation}
\begin{equation}
\label{eq:sm-S9}
\begin{aligned}
\mathcal S^{(5)}_{ij}
={}&
\mathcal A^2_{jilk\,nm}
\left(
G^{(1)}_{kl}G^{(4)}_{mn}
+
G^{(2)}_{kl}G^{(3)}_{mn}
\right)
+
\frac12
\mathcal A^3_{jilk\,nm\,qp}
\left(
G^{(1)}_{kl}G^{(1)}_{mn}G^{(3)}_{pq}
+
G^{(1)}_{kl}G^{(2)}_{mn}G^{(2)}_{pq}
\right)
\\
&+
\frac16
\mathcal A^4_{ji\,l_1k_1l_2k_2l_3k_3l_4k_4}
G^{(1)}_{k_1l_1}
G^{(1)}_{k_2l_2}
G^{(1)}_{k_3l_3}
G^{(2)}_{k_4l_4}
\\ &+ \frac1{120}
\mathcal A^5_{ji\,l_1k_1l_2k_2l_3k_3l_4k_4l_5k_5}
G^{(1)}_{k_1l_1}
G^{(1)}_{k_2l_2}
G^{(1)}_{k_3l_3}
G^{(1)}_{k_4l_4}
G^{(1)}_{k_5l_5}
\\
&-
\bar p
\left[
\mathbf G^{(1)}\mathbf G^{(4)}
+
\mathbf G^{(2)}\mathbf G^{(3)}
+
\mathbf G^{(3)}\mathbf G^{(2)}
+
\mathbf G^{(4)}\mathbf G^{(1)}
\right]_{ji}
\\ &+
\bar p
\left[
(\mathbf G^{(1)})^2\mathbf G^{(3)}
+
\mathbf G^{(1)}\mathbf G^{(3)}\mathbf G^{(1)}
+
\mathbf G^{(3)}(\mathbf G^{(1)})^2
\right]_{ji}
\\
&+
\bar p
\left[
\mathbf G^{(1)}(\mathbf G^{(2)})^2
+
\mathbf G^{(2)}\mathbf G^{(1)}\mathbf G^{(2)}
+
(\mathbf G^{(2)})^2\mathbf G^{(1)}
\right]_{ji}
\\
&-
\bar p
\left[
(\mathbf G^{(1)})^3\mathbf G^{(2)}
+
(\mathbf G^{(1)})^2\mathbf G^{(2)}\mathbf G^{(1)}
+
\mathbf G^{(1)}\mathbf G^{(2)}(\mathbf G^{(1)})^2
+
\mathbf G^{(2)}(\mathbf G^{(1)})^3
\right]_{ji}
\\
&+
\bar p
\left[
(\mathbf G^{(1)})^5
\right]_{ji}
+
p^{(1)}G^{(4)}_{ji}
+
p^{(2)}G^{(3)}_{ji}
+
p^{(3)}G^{(2)}_{ji}
+
p^{(4)}G^{(1)}_{ji}
\\ &-
p^{(1)}
\left[
\mathbf G^{(1)}\mathbf G^{(3)}
+
(\mathbf G^{(2)})^2
+
\mathbf G^{(3)}\mathbf G^{(1)}
\right]_{ji}
\\
&-
p^{(2)}
\left[
\mathbf G^{(1)}\mathbf G^{(2)}
+
\mathbf G^{(2)}\mathbf G^{(1)}
\right]_{ji}
-
p^{(3)}
\left[
(\mathbf G^{(1)})^2
\right]_{ji}
\\ &+
p^{(1)}
\left[
(\mathbf G^{(1)})^2\mathbf G^{(2)}
+
\mathbf G^{(1)}\mathbf G^{(2)}\mathbf G^{(1)}
+
\mathbf G^{(2)}(\mathbf G^{(1)})^2
\right]_{ji}
\\
&+
p^{(2)}
\left[
(\mathbf G^{(1)})^3
\right]_{ji}
-
p^{(1)}
\left[
(\mathbf G^{(1)})^4
\right]_{ji}.
\end{aligned}
\end{equation}
The order of the matrix products in Eqs. \eqref{eq:sm-S6}--\eqref{eq:sm-S9} must be retained because the gradient matrices do not commute in general.

The two-dimensional incompressibility source is already given in recursive form in Eq. \eqref{eq:main-11} of the main text. Using the same notation, its explicit forms through fifth order are
\begin{equation}
\label{eq:sm-S10}
\mathcal D^{(2)} = -\frac12 \left[ G^{(1)}_{ii}G^{(1)}_{jj} - G^{(1)}_{ij}G^{(1)}_{ji} \right],
\end{equation}
\begin{equation}
\label{eq:sm-S11}
\mathcal D^{(3)} = - \left[ G^{(1)}_{ii}G^{(2)}_{jj} - G^{(1)}_{ij}G^{(2)}_{ji} \right],
\end{equation}
\begin{equation}
\label{eq:sm-S12}
\mathcal D^{(4)} ={} - \left[ G^{(1)}_{ii}G^{(3)}_{jj} - G^{(1)}_{ij}G^{(3)}_{ji} \right] - \frac12 \left[ G^{(2)}_{ii}G^{(2)}_{jj} - G^{(2)}_{ij}G^{(2)}_{ji} \right],
\end{equation}
and
\begin{equation}
\label{eq:sm-S13}
\mathcal D^{(5)} ={} - \left[ G^{(1)}_{ii}G^{(4)}_{jj} - G^{(1)}_{ij}G^{(4)}_{ji} \right] - \left[ G^{(2)}_{ii}G^{(3)}_{jj} - G^{(2)}_{ij}G^{(3)}_{ji} \right].
\end{equation}

The nonlinear sources are assembled harmonic by harmonic. For a product of \(s\) gradient fields,
\begin{equation}
\label{eq:sm-S14}
\widehat{ \left( \mathbf G^{(a_1)} \cdots \mathbf G^{(a_s)} \right) }^{(q)} = \sum_{q_1+\cdots+q_s=q} \hat{\mathbf G}^{(a_1,q_1)} \cdots \hat{\mathbf G}^{(a_s,q_s)}.
\end{equation}
For the constitutive contribution, the same rule gives
\begin{equation}
\label{eq:sm-S15}
\widehat{ \mathcal A^{s}_{ji\,l_1k_1\cdots l_sk_s} G^{(a_1)}_{k_1l_1}\cdots G^{(a_s)}_{k_sl_s} }^{(q)}  = \sum_{q_1+\cdots+q_s=q} \mathcal A^{s}_{ji\,l_1k_1\cdots l_sk_s} \hat G^{(a_1,q_1)}_{k_1l_1}\cdots \hat G^{(a_s,q_s)}_{k_sl_s}.
\end{equation}
A term generated by the incremental pressure satisfies
\begin{equation}
\label{eq:sm-S16}
\widehat{ p^{(\ell)} \left( \mathbf G^{(a_1)} \cdots \mathbf G^{(a_s)} \right) }^{(q)}  = \sum_{q_0+q_1+\cdots+q_s=q} \hat p^{(\ell,q_0)} \hat{\mathbf G}^{(a_1,q_1)} \cdots \hat{\mathbf G}^{(a_s,q_s)}.
\end{equation}
Thus the constitutive terms and the terms proportional to \(\bar p\) satisfy \(a_1+\cdots+a_s=n\), while the terms generated by \(\tilde p\) satisfy \(\ell+a_1+\cdots+a_s=n\). In each case, the harmonic indices sum to the target wavenumber \(q\).

Equations \eqref{eq:sm-S5}, \eqref{eq:sm-S10}, and \eqref{eq:sm-S14}--\eqref{eq:sm-S16} provide the nonlinear forcing required at arbitrary perturbation order. Once the lower-order fields are known, they determine \(\hat{\mathcal S}^{(n,q)}\) and \(\hat{\mathcal D}^{(n,q)}\) for every generated harmonic.

\section{Constrained Stroh reduction and Fredholm projection}\label{constrained-stroh-reduction-and-fredholm-projection}

The incompressible Stroh system in Section 3 is obtained by eliminating the order-\(n\) pressure together with the incompressibility constraint. We use the same notation as in the main text. In particular, \(\mathbf Q_\kappa\), \(\mathbf R_\kappa\), and \(\mathbf T_\kappa\) are defined in Eq. \eqref{eq:main-18}, while \(\zeta_\kappa\) and \(\mathbf b_\kappa\) are introduced immediately before Eq. \eqref{eq:main-19}.
The residual \(\mathcal C^{(n)}\) defined in Eq. \eqref{eq:main-17} vanishes for every admissible incompressible field. The two stress relations in Eq. \eqref{eq:main-13} can therefore be replaced by the equivalent augmented relations
\begin{equation}
\label{eq:sm-S17}
\boldsymbol\chi^{(n)}_1 ={} \mathbf Q_\kappa\mathbf u^{(n)}_{,1} + \mathbf R_\kappa\mathbf u^{(n)}_{,2} - p^{(n)}\mathbf e_1 + \mathbf s^{(n)}_1 - \kappa\mathcal D^{(n)}\mathbf e_1, \qquad \boldsymbol\chi^{(n)}_2 ={} \mathbf R_\kappa^{\mathrm{T}}\mathbf u^{(n)}_{,1} + \mathbf T_\kappa\mathbf u^{(n)}_{,2} - p^{(n)}\mathbf e_2 + \mathbf s^{(n)}_2 - \kappa\mathcal D^{(n)}\mathbf e_2.
\end{equation}
Because the augmentation is proportional to \(\mathcal C^{(n)}=0\), Eq. \eqref{eq:sm-S17} is equivalent to the original constrained stress relations; \(\kappa\) is chosen so that \(\mathbf T_\kappa\) is invertible.

Using the stress potential introduced in Eq. \eqref{eq:main-14}, the second relation in Eq. \eqref{eq:sm-S17} becomes
\begin{equation}
\label{eq:sm-S18}
\boldsymbol\phi^{(n)}_{,1} = \mathbf R_\kappa^{\mathrm{T}}\mathbf u^{(n)}_{,1} + \mathbf T_\kappa\mathbf u^{(n)}_{,2} - p^{(n)}\mathbf e_2 + \mathbf s^{(n)}_2 - \kappa\mathcal D^{(n)}\mathbf e_2.
\end{equation}
Solving Eq. \eqref{eq:sm-S18} for the depth derivative gives
\begin{equation}
\label{eq:sm-S19}
\mathbf u^{(n)}_{,2} = \mathbf T_\kappa^{-1} \Big[ \boldsymbol\phi^{(n)}_{,1} - \mathbf R_\kappa^{\mathrm{T}}\mathbf u^{(n)}_{,1} + p^{(n)}\mathbf e_2  - \mathbf s^{(n)}_2 + \kappa\mathcal D^{(n)}\mathbf e_2 \Big].
\end{equation}
Substitution of Eq. \eqref{eq:sm-S19} into the exact order-\(n\) incompressibility condition gives the incremental pressure explicitly,
\begin{equation}
\label{eq:sm-S20}
\begin{aligned} p^{(n)} = \zeta_\kappa^{-1} \Big[ & (1-\kappa\zeta_\kappa)\mathcal D^{(n)} + \mathbf b_\kappa^{\mathrm{T}}\mathbf u^{(n)}_{,1}- \mathbf e_2^{\mathrm{T}} \mathbf T_\kappa^{-1} \boldsymbol\phi^{(n)}_{,1} + \mathbf e_2^{\mathrm{T}} \mathbf T_\kappa^{-1} \mathbf s^{(n)}_2 \Big]. \end{aligned}
\end{equation}
The derivative \(\boldsymbol\phi^{(n)}_{,1}\) in Eq. \eqref{eq:sm-S20} follows directly from the definition of the stress potential. The stress potential itself is not the traction. Its \(x_1\) derivative gives the transverse stress column.
Substituting Eq. \eqref{eq:sm-S20} into Eq. \eqref{eq:sm-S19} and into the first relation of Eq. \eqref{eq:sm-S17} gives the Stroh blocks and the source vector reported in Eqs. \eqref{eq:main-20} and \eqref{eq:main-23} of the main text. No additional constitutive approximation is introduced in this elimination.

The final operator is independent of \(\kappa\). This can be checked directly. Since
\begin{equation}
\label{eq:sm-S21}
\frac{d\mathbf T_\kappa^{-1}}{d\kappa} = - \mathbf T_\kappa^{-1} \mathbf e_2\mathbf e_2^{\mathrm{T}} \mathbf T_\kappa^{-1},
\end{equation}
the auxiliary quantities introduced immediately before Eq. \eqref{eq:main-19} satisfy
\begin{equation}
\label{eq:sm-S22}
\frac{d\zeta_\kappa}{d\kappa} = -\zeta_\kappa^2, \qquad \frac{d\mathbf b_\kappa^{\mathrm{T}}}{d\kappa} = -\zeta_\kappa\mathbf b_\kappa^{\mathrm{T}}.
\end{equation}
These identities give
\begin{equation}
\label{eq:sm-S23}
\frac{d\mathbf N_1}{d\kappa} = \mathbf0, \qquad \frac{d\mathbf N_2}{d\kappa} = \mathbf0, \qquad \frac{d\mathbf N_3}{d\kappa} = \mathbf0.
\end{equation}
They also give
\begin{equation}
\label{eq:sm-S24}
\frac{d}{d\kappa} \left( \zeta_\kappa^{-1} \mathbf T_\kappa^{-1}\mathbf e_2 \right) = \mathbf0, \qquad \frac{d}{d\kappa} \left( \zeta_\kappa^{-1}\mathbf b_\kappa \right) = \mathbf0.
\end{equation}
Equations \eqref{eq:sm-S23} and \eqref{eq:sm-S24} establish the \(\kappa\)-independence of \(\mathbf N\) and \(\mathbf H^{(n)}\).

We next derive the Fredholm projection used in Section 4. Consider the resonant harmonic \(m_c\) at the critical state. Let \((\hat{\mathbf u}^{(n,m_c)},\hat p^{(n,m_c)})\) denote the unknown order-\(n\) resonant correction. Let \((\hat{\mathbf u}_c,\hat p_c)\) denote the critical mode. Integration of the order-\(n\) equilibrium equation against the complex-conjugate critical displacement gives
\begin{equation}
\label{eq:sm-S25}
0 = \sum_L \int_{\mathcal I_{L,c}} \Big[ \overline{\hat G_{c,ij}}\, \bar{\mathcal A}^{1}_{jilk} \hat G^{(n,m_c)}_{kl}  + \overline{\hat G_{c,ij}}\, \hat{\mathcal S}^{(n,m_c)}_{ij} \Big]dx_2 .
\end{equation}
The term containing the unknown pressure \(\hat p^{(n,m_c)}\) vanishes because the critical mode is incompressible. Boundary terms vanish at the traction-free surface and in the far field. The two interface terms cancel because displacement and traction are continuous across the bonded interface. The periodic contributions cancel over one wavelength.

The critical equilibrium equation can be tested in the reverse order with the resonant correction. Using the major symmetry of \(\bar{\boldsymbol{\mathcal A}}^1\) gives
\begin{equation}
\label{eq:sm-S26}
\sum_L \int_{\mathcal I_{L,c}} \overline{\hat G_{c,ij}}\, \bar{\mathcal A}^{1}_{jilk} \hat G^{(n,m_c)}_{kl}\,dx_2 = \sum_L \int_{\mathcal I_{L,c}} \overline{\hat p_c}\, \hat G^{(n,m_c)}_{ii}\,dx_2 .
\end{equation}
The order-\(n\) incompressibility condition gives \(\hat G^{(n,m_c)}_{ii}=\hat{\mathcal D}^{(n,m_c)}\). Substitution of Eq. \eqref{eq:sm-S26} into Eq. \eqref{eq:sm-S25} therefore yields
\begin{equation}
\label{eq:sm-S27}
\mathcal P_{\mathrm{nl}}^{(n)} = \sum_L \int_{\mathcal I_{L,c}} \left[ \overline{\hat G_{c,ij}}\, \hat{\mathcal S}^{(n,m_c)}_{ij} + \overline{\hat p_c}\, \hat{\mathcal D}^{(n,m_c)} \right]dx_2 .
\end{equation}
Equation \eqref{eq:sm-S27} is the nonlinear contribution to the Fredholm projection used in Eq. \eqref{eq:main-37} of the main text.

The same projection applied to the slow-time inertial term gives the modal mass
\begin{equation}
\label{eq:sm-S28}
\mathcal M = \sum_L \int_{\mathcal I_{L,c}} \rho_L \overline{\hat u_{c,i}} \hat u_{c,i}\,dx_2 >0.
\end{equation}
This is the normalization factor used in \(c_0=\gamma_0/\mathcal M\), \(c_1=\gamma_1/\mathcal M\), and \(c_2^\star=\gamma_2^\star/\mathcal M^\star\). The loading projection \(\gamma_0\) requires an additional derivative with respect to the base stretch. Its construction is given in Appendix~\ref{loading-derivative-and-neo-hookean-cubic-reduction}.

\section{Loading derivative and neo-Hookean cubic reduction}\label{loading-derivative-and-neo-hookean-cubic-reduction}

The cubic amplitude equation contains two coefficients that require additional derivation. The first is the loading coefficient \(\gamma_0\), which measures the variation of the complete linear bifurcation problem with the base stretch. The second is the cubic coefficient \(\gamma_1\). For the neo-Hookean system considered in Section 5, \(\gamma_1\) admits a substantial simplification.

The material profile is fixed in \(X_2\), while the current depth coordinate and film thickness vary with \(\lambda\). We pull neighbouring states back to the critical current depth coordinate.
To compare neighbouring base states on a common domain, we use the current depth coordinate at the critical state,
\begin{equation}
\label{eq:sm-S29}
x_{2c} = \frac{X_2}{\lambda_c} = \frac{\lambda}{\lambda_c}x_2, \qquad x_2 = \frac{\lambda_c}{\lambda}x_{2c}.
\end{equation}
Thus \(x_{2c}=x_2\) at \(\lambda=\lambda_c\). The in-plane current coordinate \(x_1\) and the critical current wavenumber \(m_c\) are held fixed. At a simple critical extremum, the equivalent material-wavenumber parametrization gives the same first-order loading derivative because \(\partial\lambda_{\mathrm{bif}}/\partial m=0\) at \(m=m_c\). Define
\begin{equation}
\label{eq:sm-S30}
\Lambda = \frac{\lambda}{\lambda_c}, \qquad J_\Lambda=\Lambda^{-1}.
\end{equation}
For a field written on the fixed critical current domain, define
\begin{equation}
\label{eq:sm-S31}
\mathcal G_{i1}[\mathbf u;\lambda] = u_{i,1}, \qquad \mathcal G_{i2}[\mathbf u;\lambda] = \Lambda\,u_{i,2c}, \qquad \operatorname{div}_{\lambda}\mathbf u = u_{1,1} + \Lambda u_{2,2c},
\end{equation}
where a subscript \(2c\) denotes differentiation with respect to \(x_{2c}\). At the critical state, \(\Lambda=1\) and \(\mathcal G_{ij}=\hat G_{c,ij}\) when evaluated on the critical mode.

The mixed linear sesquilinear form on the critical current domain is then
\begin{equation}
\label{eq:sm-S32}
\mathfrak b_\lambda [(\mathbf v,q),(\mathbf u,p)] = \sum_L \int_{\mathcal I_{L,c}} J_\Lambda \Big[ \overline{\mathcal G_{ij}[\mathbf v;\lambda]}\, \bar{\mathcal A}^{1}_{jilk} (\lambda,X_2=\lambda_c x_{2c})\, \mathcal G_{kl}[\mathbf u;\lambda]  - p\,\overline{\operatorname{div}_{\lambda}\mathbf v} - \overline q\,\operatorname{div}_{\lambda}\mathbf u \Big]dx_{2c} .
\end{equation}
The loading coefficient is defined by differentiating the complete mixed sesquilinear form while the critical mode is held fixed,
\begin{equation}
\label{eq:sm-S33}
\gamma_0 = \left. \frac{\partial}{\partial\lambda} \mathfrak b_\lambda \left[ (\hat{\mathbf u}_c,\hat p_c), (\hat{\mathbf u}_c,\hat p_c) \right] \right|_{\lambda=\lambda_c}.
\end{equation}
This derivative includes the change of the current depth measure, the transformed depth derivatives, the instantaneous tensor, and the mixed incompressibility terms. Differentiating Eq. \eqref{eq:sm-S32} gives
\begin{equation}
\label{eq:sm-S34}
\begin{aligned} \gamma_0 = \sum_L \int_{\mathcal I_{L,c}} \Bigg\{& \overline{\hat G_{c,ij}} \left[ \left. \frac{\partial\bar{\mathcal A}^{1}_{jilk}} {\partial\lambda} \right|_{X_2} + \frac{\delta_{j2}+\delta_{l2}-1}{\lambda_c} \bar{\mathcal A}^{1}_{jilk} \right] \hat G_{c,kl} + \frac{1}{\lambda_c} \left( \hat p_c\,\overline{\hat G_{c,11}} + \overline{\hat p_c}\,\hat G_{c,11} \right) \Bigg\}dx_{2c} . \end{aligned}
\end{equation}
All quantities without an explicit \(\lambda\)-derivative are evaluated at the critical state. Equation \eqref{eq:sm-S34} is the loading sensitivity used in Eq. \eqref{eq:main-38} of the main text. It is not equal to the derivative of \(\bar{\mathcal A}^{1}\) alone.

For the incompressible neo-Hookean material used in Section 5, let \(\bar{\mathbf B}=\bar{\mathbf F}\bar{\mathbf F}^{\mathrm T}\) denote the base-state left Cauchy--Green tensor. Then
\begin{equation}
\label{eq:sm-S35}
\bar{\mathcal A}^{1}_{jilk} = \mu(X_2) \left( \delta_{ik}\bar B_{jl} + \lambda^{-2}\delta_{jk}\delta_{il} \right),
\end{equation}
and
\begin{equation}
\label{eq:sm-S36}
\left. \frac{\partial\bar{\mathcal A}^{1}_{jilk}} {\partial\lambda} \right|_{X_2} = \mu(X_2) \left[ \delta_{ik} \frac{d\bar B_{jl}}{d\lambda} - 2\lambda^{-3}\delta_{jk}\delta_{il} \right].
\end{equation}
Using the first-order incompressibility condition, Eq. \eqref{eq:sm-S34} reduces to
\begin{equation}
\label{eq:sm-S37}
\begin{aligned} \gamma_0 = \sum_L \int_{\mathcal I_{L,c}} \Bigg\{& \mu_L \left[ \lambda_c \sum_i |\hat G_{c,i1}|^2 - \lambda_c^{-3} \sum_i |\hat G_{c,i2}|^2  - 2\lambda_c^{-3} \sum_{i,j} \overline{\hat G_{c,ij}} \hat G_{c,ji} \right] \\ &+ \frac{1}{\lambda_c} \left( \hat p_c\,\overline{\hat G_{c,11}} + \overline{\hat p_c}\,\hat G_{c,11} \right) \Bigg\}dx_{2c} . \end{aligned}
\end{equation}
Here \(\mu_L\) denotes the shear modulus evaluated at \(X_2=\lambda_c x_{2c}\) in each layer. It is constant in the film and may vary with depth in the substrate.

We next simplify the cubic nonlinear projection. For the neo-Hookean energy, the strain-energy density is quadratic in \(\mathbf F\). Hence
\begin{equation}
\label{eq:sm-S38}
\mathcal A^s=0, \qquad s\ge2.
\end{equation}
The cubic source contains only the pressure and incompressibility contributions generated by lower-order fields. For compactness in the remainder of this section, define
\begin{equation}
\label{eq:sm-S39}
\mathbf G_c = \hat{\mathbf G}_c, \qquad \mathbf G_0 = \hat{\mathbf G}^{(2,0)}, \qquad \mathbf G_2 = \hat{\mathbf G}^{(2,2m_c)},
\end{equation}
and
\begin{equation}
\label{eq:sm-S40}
\pi_c=\hat p_c, \qquad \pi_0=\hat p^{(2,0)}, \qquad \pi_2=\hat p^{(2,2m_c)}.
\end{equation}
All coefficient fields below are defined after the associated amplitude factors have been extracted, consistently with Section 4.2. The subscripts \(0\) and \(2\) in Eq. \eqref{eq:sm-S39} refer to the mean and second harmonics. They do not denote perturbation order.

First-order incompressibility gives \(\operatorname{tr}\mathbf G_c=0\). At second order, the mean and second-harmonic constraints give
\begin{equation}
\label{eq:sm-S41}
\operatorname{tr}\mathbf G_0 = \operatorname{tr} (\mathbf G_c\overline{\mathbf G_c}), \qquad \operatorname{tr}\mathbf G_2 = \frac12 \operatorname{tr} (\mathbf G_c\mathbf G_c).
\end{equation}
For any trace-free \(2\times2\) matrix \(\mathbf A\),
\begin{equation}
\label{eq:sm-S42}
\mathbf A\mathbf A = \frac12 \operatorname{tr}(\mathbf A\mathbf A)\mathbf I.
\end{equation}
A second useful identity is
\begin{equation}
\label{eq:sm-S43}
\mathbf A\mathbf B+\mathbf B\mathbf A = \operatorname{tr}(\mathbf B)\mathbf A + \operatorname{tr}(\mathbf A\mathbf B)\mathbf I, \qquad \operatorname{tr}\mathbf A=0.
\end{equation}
Equations \eqref{eq:sm-S41}--\eqref{eq:sm-S43} reduce the resonant cubic pressure terms. In particular,
\begin{equation}
\label{eq:sm-S44}
\mathbf G_c\mathbf G_c\overline{\mathbf G_c} + \mathbf G_c\overline{\mathbf G_c}\mathbf G_c + \overline{\mathbf G_c}\mathbf G_c\mathbf G_c = \operatorname{tr} (\mathbf G_c\overline{\mathbf G_c})\mathbf G_c + \frac12 \operatorname{tr} (\mathbf G_c\mathbf G_c) \overline{\mathbf G_c}.
\end{equation}
The non-isotropic parts of the base-pressure terms therefore cancel against the corresponding second-order trace contributions. The remaining isotropic terms vanish when contracted with the trace-free critical gradient. The terms proportional to \(p^{(1)}\) that contain two first-order gradients vanish for the same reason.

The resonant third-order incompressibility source reduces to
\begin{equation}
\label{eq:sm-S45}
\hat{\mathcal D}^{(3,m_c)} = \operatorname{tr} (\mathbf G_c\mathbf G_0) + \operatorname{tr} (\overline{\mathbf G_c}\mathbf G_2).
\end{equation}
Substitution into the Fredholm projection in Eq. \eqref{eq:sm-S27} gives
\begin{equation}
\label{eq:sm-S46}
\begin{aligned} \gamma_1 = \sum_L \int_{\mathcal I_{L,c}} \Big\{& \overline{G_{c,ij}} \left( \pi_c G_{0,ji} + \overline{\pi_c}\,G_{2,ji} + \pi_0 G_{c,ji} + \pi_2\overline{G_{c,ji}} \right) + \overline{\pi_c} \left( G_{c,ij}G_{0,ji} + \overline{G_{c,ij}}G_{2,ji} \right) \Big\}dx_2 . \end{aligned}
\end{equation}
Equation \eqref{eq:sm-S46} is the neo-Hookean semi-analytical cubic reduction used in Section 5 of the main text. The fields \(\mathbf G_c\), \(\mathbf G_0\), and \(\mathbf G_2\) are obtained from the linear critical problem and the two second-order forced problems. No third-order boundary-value problem is required to evaluate
\begin{equation}
\label{eq:sm-S47}
c_1=\frac{\gamma_1}{\mathcal M}.
\end{equation}

The reduction in Eq. \eqref{eq:sm-S46} is specific to two-dimensional incompressible neo-Hookean elasticity. It follows from the vanishing higher constitutive tensors together with the trace identities in Eqs. \eqref{eq:sm-S41}--\eqref{eq:sm-S44}. The static reduced equation inherits the real hyperelastic potential and translational phase symmetry; its scalar coefficients, including \(\gamma_1\) and the quintic coefficient at cubic degeneracy, are therefore real.

\section{Fifth-order solvability at cubic degeneracy}\label{fifth-order-solvability-at-cubic-degeneracy}

This section gives the technical details needed for the fifth-order construction in Section 4.3 of the main text. We consider a simple cubic-degenerate system \(\boldsymbol\xi^\star\in\Sigma_3\), with \(c_1^\star=0\) and \(\nabla_{\boldsymbol\xi}c_1(\boldsymbol\xi^\star)\neq\mathbf0\). All starred quantities are evaluated at this system. The required harmonic sets are \(\mathcal Q_1=\{\pm m_c^\star\}\), \(\mathcal Q_2=\{0,\pm2m_c^\star\}\), \(\mathcal Q_3=\{\pm m_c^\star,\pm3m_c^\star\}\), \(\mathcal Q_4=\{0,\pm2m_c^\star,\pm4m_c^\star\}\), and \(\mathcal Q_5=\{\pm m_c^\star,\pm3m_c^\star,\pm5m_c^\star\}\). At cubic degeneracy, the third-order resonant forcing does not vanish in general. Only its Fredholm projection vanishes. The \(m_c^\star\)-harmonic therefore satisfies
\begin{equation}
\label{eq:sm-S48}
\mathcal L_{m_c^\star,c}^\star \hat{\boldsymbol\eta}^{(3,m_c^\star)} = \hat{\mathbf H}^{(3,m_c^\star)},
\end{equation}
together with the compatibility condition obtained from Eq. \eqref{eq:main-37} of the main text. Since \(c_1^\star=0\), this compatibility condition is satisfied. Equation \eqref{eq:sm-S48} must still be solved because its particular solution enters the fifth-order forcing.

The critical operator has a one-dimensional null space. Hence
\begin{equation}
\label{eq:sm-S49}
\hat{\boldsymbol\eta}^{(3,m_c^\star)} = \hat{\boldsymbol\eta}^{(3,m_c^\star)}_p + \vartheta \hat{\boldsymbol\eta}_c^\star.
\end{equation}
The gauge condition in Eq. \eqref{eq:main-43} of the main text fixes the coefficient \(\vartheta\). For an arbitrary particular solution,
\begin{equation}
\label{eq:sm-S50}
\vartheta = - \frac{ \displaystyle \sum_L \int_{\mathcal I_{L,c}^\star} \rho_L \overline{\hat{\mathbf u}_c^\star} \cdot \hat{\mathbf u}^{(3,m_c^\star)}_p \,dx_2 }{ \mathcal M^\star }.
\end{equation}

A change in the critical-mode component of the third-order field changes the amplitude coordinate at cubic order. For steady real amplitudes, the two coordinates have the form \(a=b+\vartheta b^3+O(b^5)\). At exact cubic degeneracy, \(\Delta=O(b^4)\), and substitution into the retained steady equation gives
\begin{equation}
\label{eq:sm-gauge-transformed-amplitude-equation}
-c_0^\star\Delta a+c_2^\star a^5
=
-c_0^\star\Delta b+c_2^\star b^5+O(b^7).
\end{equation}
The contribution \(-c_0^\star\vartheta\Delta b^3\) and the correction to \(a^5\) both enter at order \(b^7\). The quintic coefficient at \(c_1^\star=0\) is therefore unchanged by this third-order gauge freedom. This statement assumes a fixed normalization of the leading critical mode; a separate rescaling of that mode changes coefficient magnitudes but not their signs.

The fifth-order resonant source is assembled from the lower-order fields using the recursive formulas in Appendix~\ref{higher-order-nonlinear-sources}. Only combinations whose perturbation orders sum to five and whose harmonic wavenumbers sum to \(m_c^\star\) contribute to \(\hat{\mathcal S}^{(5,m_c^\star)}\) and \(\hat{\mathcal D}^{(5,m_c^\star)}\). Together with the first- and second-order fields, the additional solves are the third-order \(m_c^\star\) and \(3m_c^\star\) harmonics and the fourth-order mean, \(2m_c^\star\), and \(4m_c^\star\) harmonics. Their fifth-order resonant projection defines \(\gamma_2^\star\) in Eq. \eqref{eq:main-44}.
For the neo-Hookean material used in Section 5, \(\mathcal A^s=0\) for \(s\ge2\). The fifth-order source is generated entirely by the pressure expansion and the nonlinear incompressibility constraint. The required terms follow directly from Eqs. \eqref{eq:sm-S9} and \eqref{eq:sm-S13} of Appendix~\ref{higher-order-nonlinear-sources} after extraction of the \(m_c^\star\) harmonic.

The quintic coefficient is evaluated at the exact cubic-degenerate critical state. The material and structural parameters are held fixed during this reduction, while the load departure obeys \(\Delta=O(\varepsilon^4)\). If the resulting \(c_2^\star\) is non-zero, the solvability condition closes the leading branch at fifth order. If it also vanishes, the retained hierarchy must be extended before a nonlinear branch direction can be assigned.

\section*{Declaration of generative AI and AI-assisted technologies in the writing process}
During the preparation of this work, the authors used OpenAI's ChatGPT and Codex to assist with English-language editing, the organization of selected passages, and the development and debugging of parts of the numerical code. All AI-assisted text and code were reviewed, tested, and edited by the authors, who take full responsibility for the content of the published article.
\begingroup
\raggedright

\endgroup

\end{document}